\documentclass[11pt]{article}
\usepackage[utf8]{inputenc}
\usepackage{graphicx}
\usepackage{amsmath}
\usepackage{amssymb}
\usepackage{tabularx}
\usepackage[dvipsnames]{xcolor}
\usepackage{mwe}
\usepackage{wrapfig}
\usepackage{multirow}
\usepackage{float}
\usepackage[hyphens]{url}
\usepackage{hyperref}

\RequirePackage[left=1in,%
                right=1in,%
                top=1in,%
                bottom=1in,%
                headheight=12pt,%
                letterpaper]{geometry}%

\title{Tutorial: Circuit-based Electromagnetic Transient Simulation}
\author{Amritanshu Pandey, University of Vermont}

\graphicspath{{Figures/}}
\usepackage[font=scriptsize]{caption}

\usepackage[style=ieee, backend=biber]{biblatex}
\begin{document}

\maketitle 

\begin{table*}[htpb]
\small
\caption{Notations for Parameters and Symbols}
\centering
\begin{tabular}
{c|l}
\hline
  \textbf{Parameters} & \textbf{Description} \\
  \hline
  \multicolumn{2}{c}{Simulation Variables and Parameters.}\\
  \hline
  $Y$ & Nodal admittance matrix \\
  $V$ & Solution vector of system states \\
  $J$ & Independent source vector \\
  $k$ & iteration count \\
  $\Delta t$ & time-step \\
  $t$ & time \\
  \hline
  \multicolumn{2}{c}{DQ Transformation}\\
  \hline
  $\theta$ & Arbitrary value for stationary, synchronous, or rotating reference frame $(rad)$
  \\
  $\beta$ & Rotor angular position wrt to reference frame $(rad)$ \\
  $ \lambda$ & $2pi/3$ $(rad)$ \\
  \hline
  \multicolumn{2}{c}{Linear Circuit Example} \\
  \hline
  $R_a, R_b, R_c$ & Line resistances for phases $abc$ $(\Omega)$ \\
  $L_a, L_b, L_c$ & Line inductances for phases $abc$ $(H)$ \\
  $R^L_a, R^L_b, R^L_c$ & Load resistances for phases $abc$ $(\Omega)$ \\
  $L^L_a, L^L_b, L^L_c$ & Load inductances for phases $abc$ $(H)$ \\
  \hline
  \multicolumn{2}{c}{Induction Motor Example} \\
  \hline
  $F_{dr}$ & Nonlinear KVL equation for IM's $dr$ subcircuit \\
  $F_{ds}$ & Nonlinear KVL equation for IM's $ds$ subcircuit \\
  $F_{qr}$ & Nonlinear KVL equation for IM's $qr$ subcircuit \\
  $F_{qs}$ & Nonlinear KVL equation for IM's $qs$ subcircuit \\
  $R_s$ & Rotor resistance ($\Omega$) \\
  $R_s$ & Stator resistance ($\Omega$) \\
  $L_s$ & Stator self-inductance $(H)$ \\
  $L_r$ & Rotor self-inductance $(H)$ \\
  $L_m$ & Magnetizing inductance $(H)$ \\
  $L_{ls}$ & Stator leakage inductance $(H)$ \\
  $L_{lr}$ & Rotor leakage inductance $(H)$ \\
  $\omega_s$ & Synchronous angular velocity $(rad/s)$ \\
  $\omega_r$ & Angular velocity of the rotor $(rad/s)$ \\
  $I_{dr}$ & Current in rotor's direct axis $(A)$ \\
  $I_{ds}$ & Current in stator's direct axis $(A)$ \\
  $I_{qr}$ & Current in rotor's quadrature axis $(A)$ \\
  $I_{qs}$ & Current in stator's quadrature axis $(A)$ \\
  $V_{dr}$ & Voltage across rotor's direct axis $(V)$ \\
  $V_{ds}$ & Voltage across stator's direct axis $(V)$ \\
  $V_{qr}$ & Voltage across rotor's quadrature axis $(V)$ \\
  $V_{qs}$ & Voltage across stator's quadrature axis $(V)$ \\
  $\psi_{dr}$ & Flux linkages across rotor's direct axis $(Weber)$ \\
  $\psi_{ds}$ & Flux linkages across stator's direct axis $(Weber)$ \\
  $\psi_{qr}$ & Flux linkages across rotor's quadrature axis $(Weber)$\\
  $\psi_{qs}$ & Flux linkages across stator's quadrature axis $(Weber)$\\
  $J$ & Combined inertia of motor and load ($kgm^2$)\\
  $D$ & Combined viscous friction of motor and load $(N.m/(rad/s))$ \\
  $\theta_r$ & Motor mechanical angular position $(rad)$ \\
  $T_L$ & Motor load torque $(Nm)$ \\
  $T_E$ & Motor electrical torque $(Nm)$ \\
  $N_p$ & number of poles \\
\hline
\end{tabular}
\end{table*}

\section*{Abstract}
\noindent The growing penetration of inverter-based resources and associated controls necessitates system-wide electromagnetic transient (EMT) analyses. 
EMT tools and methods today were not designed for the scale of these analyses. 
In light of the emerging need, there is a great deal of interest in developing new techniques for fast and accurate EMT simulations for large power grids; 
the foundations of which will be built on current tools and methods. 
However, we find that educational texts covering the fundamentals and inner workings of current EMT tools are limited.
As such, there is a lack of introductory material for students and professionals interested in researching the field.
To that end, in this tutorial, we introduce the principles of EMT analyses from the circuit-theoretic viewpoint, mimicking how time-domain analyses are performed in circuit simulation tools like SPICE and Cadence. 
We perform EMT simulations for two examples, one linear and one nonlinear, including induction motor (IM) from the first principles.
By the document's end, we anticipate the readers will have a \textit{basic} understanding of how power grid EMT tools work.

\section*{Introduction}

This document is a tutorial to help students and working professionals learn the inner workings of tools that perform electromagnetic transient (EMT) simulations. The tutorial uses circuit-theory based approach to EMT simulation. The tutorial is introductory in that it shows the workings through two simple examples and expects the reader to build expertise through more detailed readings. Nonetheless, this document is an excellent first step for anyone interested in the subject, especially those interested in building their own tools. 

Before we dive into EMT models and simulation, it is pertinent to discuss and differentiate between the two types of commonly performed time-domain simulations in the power community:

\begin{itemize}
    \item \textbf{Transient Stability (TS):} Performs time-domain simulations on balanced networks (only positive sequence components). It does not include network impedance transient response and models them moving from one steady state to another instantaneously. Therefore transient stability frameworks model the network constraints with algebraic equations and only model the injection components and controls with differential algebraic equations. TS is also sometimes called root-mean-squared (RMS) transient analysis. These have been traditionally the workhorse for system-wide time-domain power grid analysis.
    \item \textbf{Electromagnetic Transient Simulation:} Works exactly as circuit-simulation tools (for instance, SPICE and CADENCE). It models all three phases, including the transient response for the network impedance components. The industry demand for large-scale EMTs is growing rapidly due to the recent events involving power electronics devices.
\end{itemize}

This document focuses on electromagnetic transient simulation. These are becoming increasingly common (with significant active research) due to the introduction of inverter-based resources on the grid. The document provides a step-by-step tutorial on performing an EMT simulation for simple networks from the first principles. It uses concepts from circuit simulation, Newton-Raphson, and numerical integration with difference methods. Good references to brush up on these topics are \cite{circuit_simulation_pileggi},\cite{newton}, and \cite{numerical_integration}. The document does not discuss modeling controls in EMT but provides sufficient background such that it can be tackled as the next step.

Furthermore, this document uses modified nodal analysis (MNA) and sometimes loop analysis to encapsulate network physics. Other alternate approaches like Tree Link Analysis (TLA) and Sparse Tableau Analysis (STA) \cite{circuit_simulation_pileggi} exist but are not covered in this document. In naive nodal analysis, the currents at each node are summed to zero (i.e., Kirchhoff's current law) to satisfy network physics. However, the method fails when the network includes voltage sources, as the current through the voltage source is not implicitly known. Therefore, to include voltage sources, the modified version adds one additional constraint per voltage source to the set of nodal equations. These additional constraints give us the currents through the voltage sources as new variables. The term \textit{modified} in MNA refers to this modification. Finally, the readers must note that anytime in this document when they come across the following terms: system matrix, $Y$ matrix, solution matrix, or simple nodal matrix, these all refer to the same thing: a set of nodal or loop equations (or sometimes a mix) for the linearized network that is being evaluated at a given time-step \textit{t}.

\section{Equivalent Circuit Approach for EMT Simulations}

We will use an equivalent circuit framework \cite{pandey2016unified}, \cite{pandey2018robust} for power grids to develop EMT models for two simple toy networks. These toy examples are based on a simple 2-bus three-phase power network with different load configurations. We consider the following load scenarios:
\begin{itemize}
    \item a linear wye-connected series RL load (see Fig. \ref{fig:ideal_voltage_network_RL_node})
    \item a non-linear induction motor load (see Fig. \ref{fig:IM_load})
\end{itemize}

\noindent Note: For both examples, we assume that the load is connected in a wye configuration and all the neutrals in the network are grounded. 

\begin{figure}[htp]
    \centering
    \includegraphics[width=12cm]{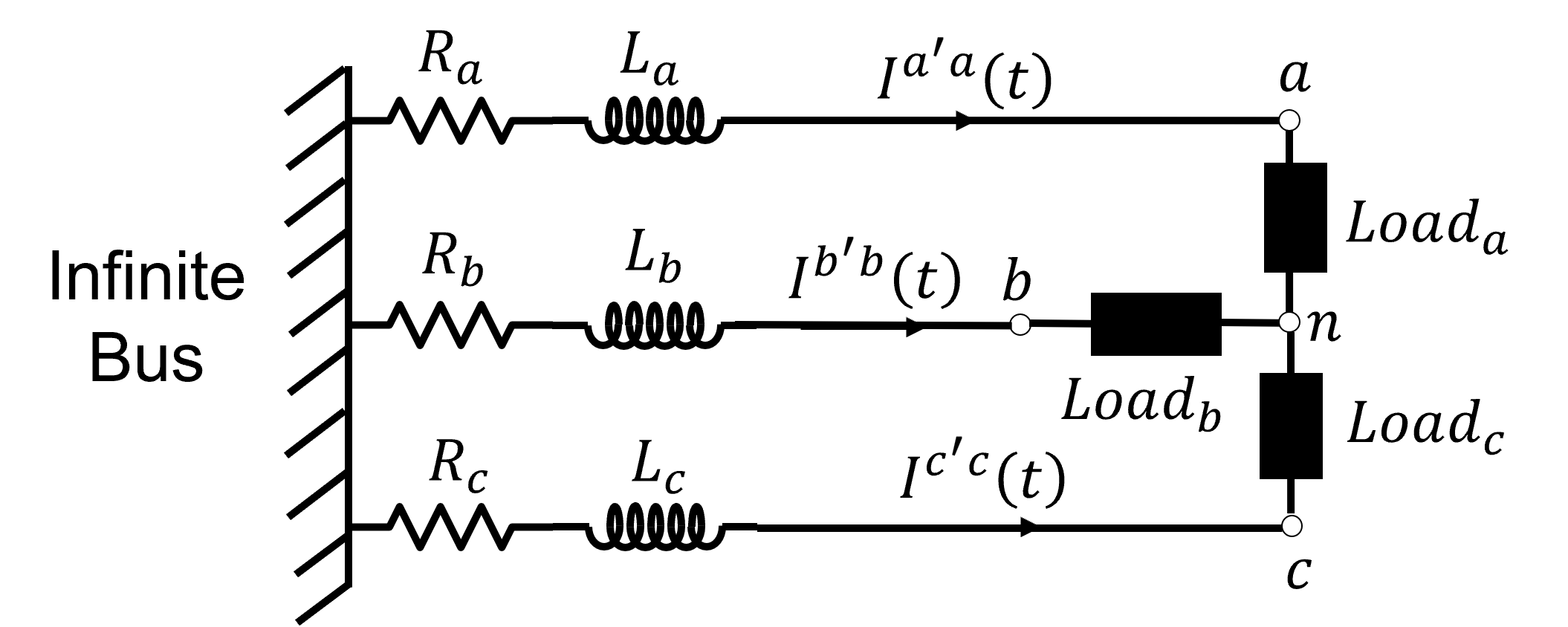}
    \caption{Power system representation of the 2-bus power network.}
    \label{fig:infinite_bus}
\end{figure}

The illustration of the two-bus power network with a generic load model is shown in Fig. \ref{fig:infinite_bus}. It includes power system verbiage such as infinite bus and wye-connected load. For the moment, the infinite bus (at nodes $a', b', c'$) can be considered the power source, and the load bus (at nodes $a, b, c$) can be considered the power sink.

%\begin{wrapfigure}{L}{0.5\textwidth}
%\end{wrapfigure} 

\begin{figure}[H]
    \centering
    \includegraphics[width=14cm]{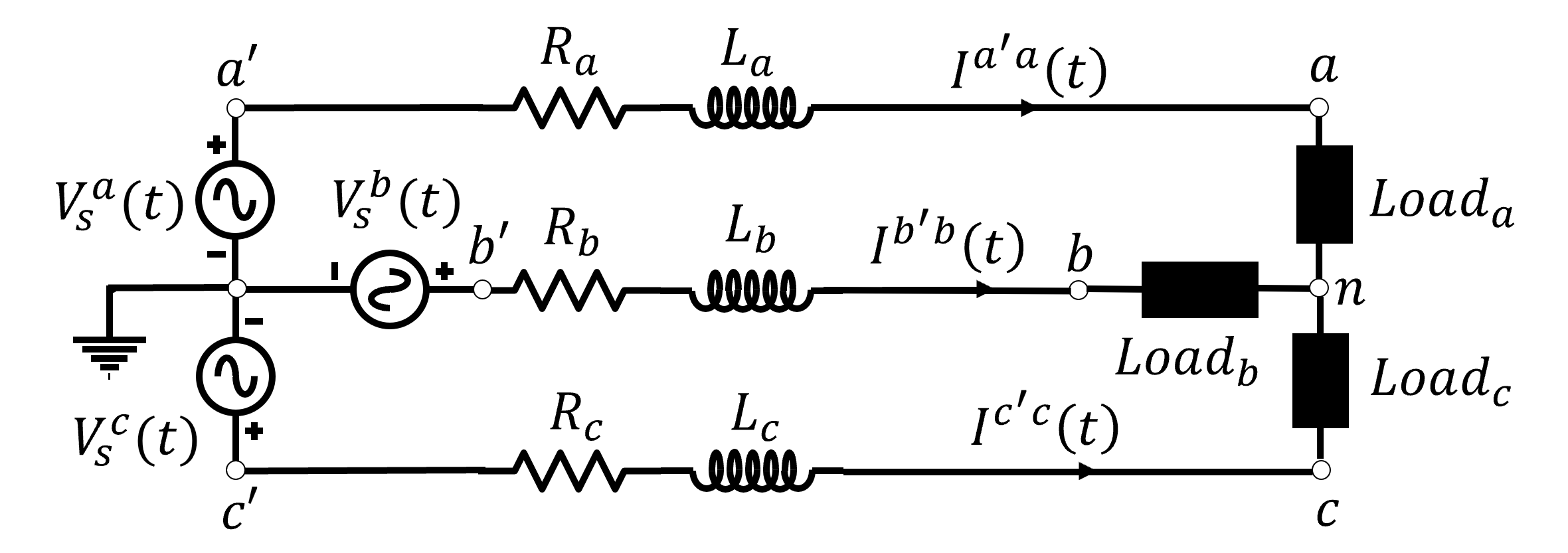}
    \caption{Power system representation of the 2-bus power network.}
    \label{fig:ideal_voltage_network}
\end{figure}

The illustration in Fig. \ref{fig:infinite_bus} is a power system representation. To perform EMT simulation with an equivalent circuit framework, as a first step, we represent any power grid-specific elements as electric circuit elements. The network in Fig. \ref{fig:infinite_bus} has three components i) an infinite bus, ii) a transmission line (a very simplified model), and iii) a load. Two of these three components (infinite bus and transmission line) are shared across the two examples. Therefore, we begin with them. The infinite bus is trivial to model as a circuit element. It is simply an ideal voltage source with a known magnitude and phase angle. Further, the frequency of this source is assumed to be either 60 Hz or 50 Hz depending on the system. For a three-phase network, the infinite bus is modeled by a set of three independent voltage sources (offset by $120^{\circ}$) connected in a wye formation. The network with an infinite bus replaced with circuit elements is shown in Fig. \ref{fig:ideal_voltage_network}. The transmission line model with R and L values already represents a set of circuit elements and does not require translation. Note that this transmission line model is greatly simplified to improve the readability of this tutorial. A more detailed model (see \cite{dommel1996emtp}), at a minimum, should include shunt capacitances and mutual impedances. Longer lines should use lumped models. Still, the readers will learn to model mutual impedances while developing the EMT model for an induction motor.

%\begin{wrapfigure}{L}{0.5\textwidth}
%\end{wrapfigure} 

The last step is to convert the load component in Fig. \ref{fig:ideal_voltage_network} into an equivalent circuit before getting started with EMT-related steps. Loads in power systems correspond to various devices ranging from large induction motors to resistive heating loads. We will translate the two load models into equivalent circuits in their respective sub-sections.

\subsection{Wye-connected RL Load}
\begin{figure}[htp]
    \centering
    \includegraphics[width=14cm]{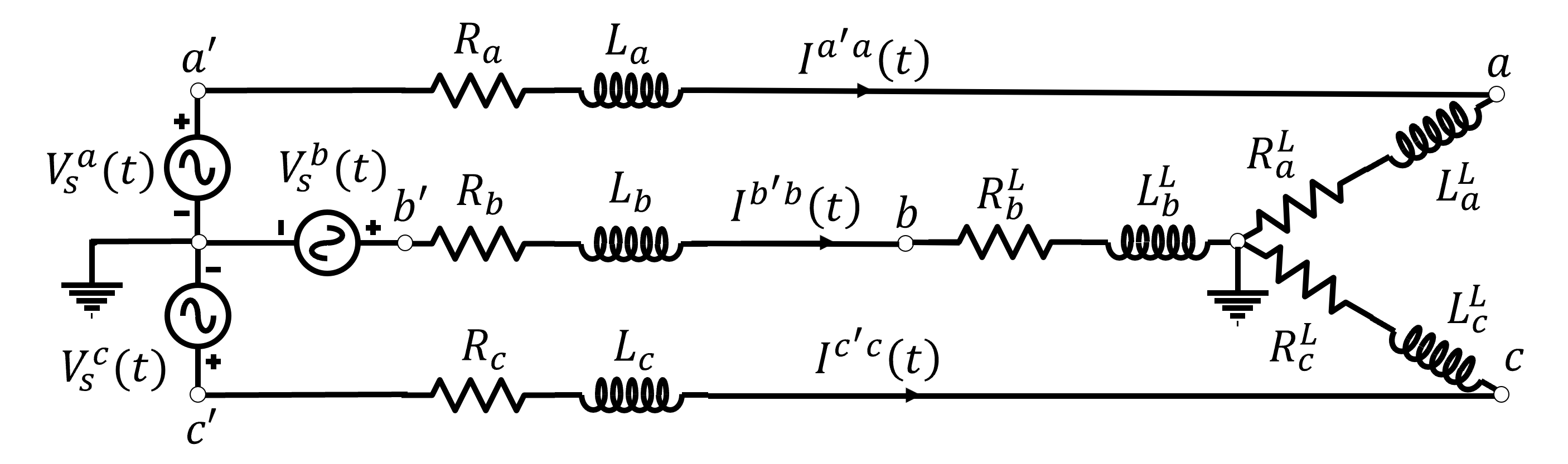}
    \caption{Power system representation of the 2-bus power network.}
    \label{fig:ideal_voltage_network_RL_node}
\end{figure}

The first example represents the load via a wye-connected series RL impedance. With series RL impedance as the load component, the equivalent circuit representation of the 2-bus network is trivial and is shown in Fig. \ref{fig:ideal_voltage_network_RL_node}. Simple observation confirms that all elements in this equivalent circuit are linear; therefore, to obtain the transient response, we only need to \textit{recursively} solve the network as a function of time. \textit{Iterative} solves with Newton-Raphson (NR) are only necessary in presence of nonlinear components. We will learn about that in the induction motor (IM) example.

%\begin{wrapfigure}{L}{0.5\textwidth}
%\end{wrapfigure} 

\noindent In current EMT tools (for instance, EMTP-RV or PSCAD), a sequence of steps that are followed  to obtain the time-domain response of a linear power network are as follows:

\begin{enumerate}
    \item First, ordinary differential equations (ODEs) corresponding to the sum of currents at each node in the network are defined
    \item Second, numerical integration approximation (such as trapezoidal or backward Euler) is applied to these ODE equations to convert them from differential algebraic equation (DAE) form to purely algebraic equation (AE) form
    \item Third, if the network is linear, the final solution for each time instance is obtained via a recursive linear solve of the algebraic equations for each time-tick
\end{enumerate}

Note that the equations for subsequent time steps are coupled through terms obtained from numerical integration. So currents at $t=n+1$ will be functions of currents at $t=n$. We will see this in more detail while constructing the companion models for various memory elements. Also, note that linear equations are recursively solved to traverse through simulation time by choosing a \textbf{time-step $(\Delta t)$}. The value of this time step depends on the local truncation error (LTE). More details on the calculation of LTE can be found in \cite{circuit_simulation_pileggi}. The general idea is to maintain a sufficient tradeoff between simulation speed (by taking larger time steps) and accuracy (by taking smaller time steps).

Now we will learn how to solve the EMT problem from the viewpoint of the equivalent circuit approach (ECA). To obtain the time-domain response of the network with ECA, we apply \textbf{two (2) circuit-simulation \textit{tricks}} to enable efficient construction of the linearized nodal equations $(Y_tV_t-J_t=0)$ for each time step $(t)$ in the transient simulation. 

\textbf{First, rather than} defining the ordinary differential equation (ODE) that captures the KCL constraint (sum of currents) at each node, we construct specialized symbolic stamps for each component in the network (e.g., resistance, inductance, etc.). Note that a stamp(s) here refers to term(s) that are to be added to the solution matrix. We use these stamps to implicitly construct the system matrix at each time tick. To do so, we parse through each circuit element in the network (for each time tick), and we add the stamps corresponding to this element in the solution matrix. Once completed, the matrix (completed by adding stamps from each component) will be equivalent to if we constructed the matrix by adding linearized nodal equations in each row. In this approach, rather than arduously constructing nodal equations for each node (imagine a circuit with million nodes), we only have to symbolically construct stamps for the number of types of circuit elements. Once that is done, all we need to do is add the stamps for each circuit element into the solution matrix.

\textbf{Second circuit simulation trick} pertains to simplifying the use of numerical integration of differential terms.
Instead of applying numerical integration to all nodal ODE equations explicitly, we develop and use companion circuits for time derivative memory elements (e.g., inductance and capacitance) in the network. By applying these two tricks, the ECA approach, rather than tackling one nodal or loop equation at a time, builds modular circuits for various elements with embedded implicit symbolic linearization and discretization. \textbf{This allows us to focus on a limited number of types of power grid elements (generator, load, etc.) instead of millions of nodal or loop equations.}

In ECA, with the two tricks, we follow the following steps sequentially to enable modular construction and solution of the system matrix at each time tick without ever having to define the complete set of nodal equations explicitly:

\begin{enumerate}
    \item Represent time-derivative elements in the circuit (e.g., inductor, capacitor etc.) with their companion circuits
    \item Develop a specialized stamp for each element in the circuit (with derivative elements replaced by their companion circuits) 
    \item $\forall t\in \tau=\{t_0,...,t_{final}\}$, parse through the overall network (generally a netlist or input file of sorts) and stamp the terms corresponding to each element in a sparse linear matrix (remember at this point, time-derivative terms are replaced by their companion models)
    \item Solve the linear problem $Y_tV_t=J_t$, to obtain the solution of state variables at time $t$, and update and repeat until final time step
\end{enumerate}

This approach allows us to implicitly obtain and solve the linearized nodal equations without constructing them symbolically. This is highly beneficial for developing heuristics and for solving large-scale networks. More on that later.

\subsubsection{Companion Model for Inductors and Trapezoidal Rule}

Returning to the RL-load example, the inductor is the only element with a time derivative term in this network (see Fig. \ref{fig:ideal_voltage_network_RL_node}). Therefore all we have to do is construct a companion model for the inductor. We then replace all the inductors in the network with their companion models. We now discuss the construction of these companion models.

This tutorial will use trapezoidal integration approximation to construct companion models. Other Euler-based or higher-order numerical integration methods can also be used following the same approach.
Applying the trapezoidal integration approximation, the companion circuit for the inductor can be derived as follows (for more details, read \cite{circuit_simulation_pileggi}):

\addtolength{\jot}{1em}
\begin{align} 
    L\frac{di}{dt} = v\\
    di = \frac{vdt}{L}\\
    \int_{t_n}^{t_{n+1}}di =  \int_{t_n}^{t_{n+1}}\frac{vdt}{L}\\
    \left(i(t_{n+1}) - i(t_n) \right) = \left(v(t_{n+1}) + v(t_n)\right)\frac{\Delta t}{2L}\\
    i(t_{n+1}) = i(t_n) + v(t_n)\frac{\Delta t}{2L} + v(t_{n+1})\frac{\Delta t}{2L} \label{eq:ind}
\end{align}

\begin{figure}[htp]
    \centering
    \includegraphics[width=10cm]{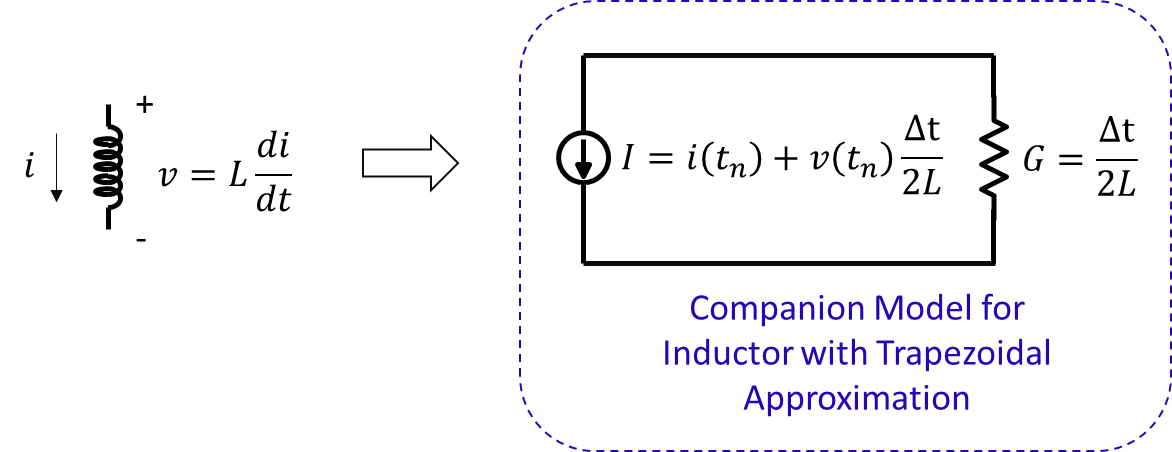}
    \caption{Inductor Trapezoidal Companion Circuit. Reconstructed from Pileggi, Carnegie Mellon ECE 18-762 Notes \cite{circuit_simulation_pileggi}.}
    \label{fig:inductor_companion_circuit}
\end{figure}

\noindent The equation in \eqref{eq:ind} has three terms. The last term $\frac{\Delta t}{2L}$ describes the relationship between current and voltage at time $t_{n+1}$, across the same branch, and is a conductance $G$. The first two terms are constants known from the states at prior time-point $t_n$ and therefore are represented via a constant current source $I$. Aggregating these elements into a circuit, we get the equivalent circuit in Fig. \ref{fig:inductor_companion_circuit}, and it is a companion model for an inductor with trapezoidal numerical integration approximation.

%\begin{wrapfigure}{L}{0.5\textwidth}
%\end{wrapfigure} 

\begin{figure}[htp]
    \centering
    \includegraphics[width=14cm]{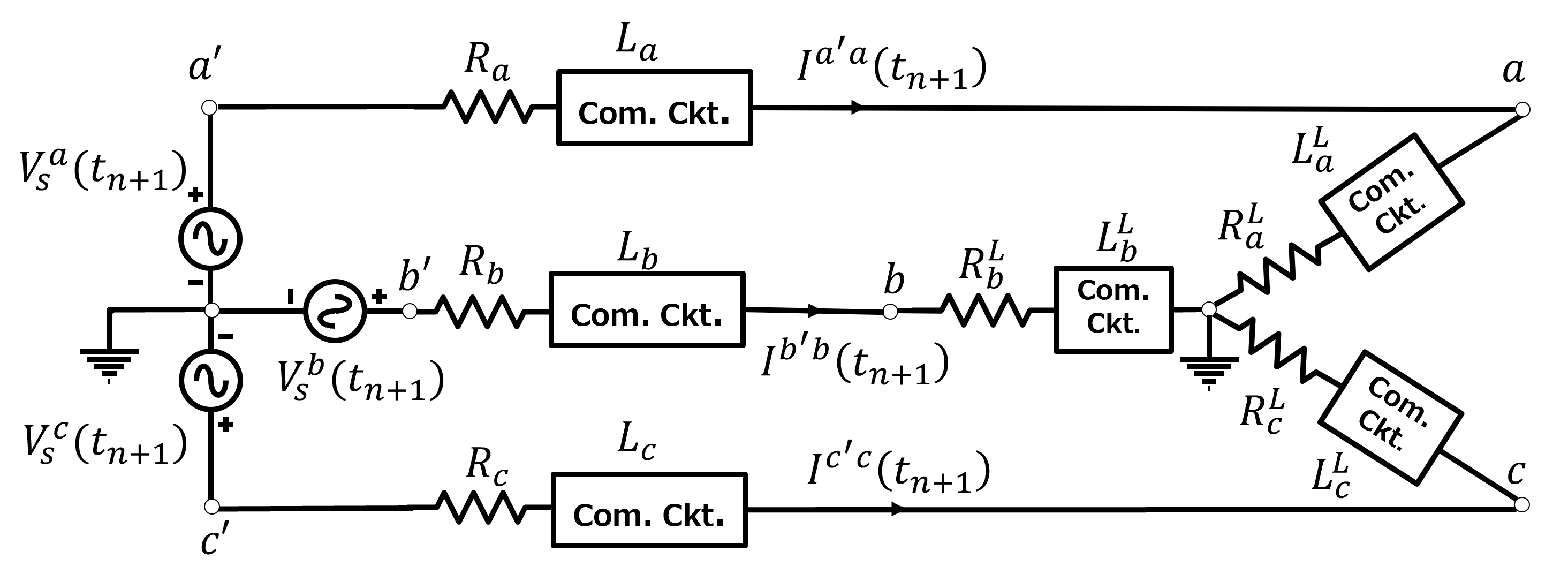}
    \caption{Equivalent circuit with companion circuits (com. ckt.).}
    \label{fig:equivalent_circuit_w_companion}
\end{figure}

With the companion model constructed, we can replace the inductor elements in Fig. \ref{fig:ideal_voltage_network_RL_node} with the respective companion circuits. Following this step, we obtain a time-dependent equivalent circuit (see Fig. \ref{fig:equivalent_circuit_w_companion}), which can be characterized entirely by linear algebraic equations. This set of equations is repeatedly solved to find the electromagnetic transient response of the network.

%\begin{wrapfigure}{L}{0.5\textwidth}
%\end{wrapfigure} 

However, as discussed, the trick is to avoid explicitly constructing the complete equivalent circuit and corresponding nodal equations $(YV - J)$ for each node. Instead, we will follow a modular approach. As we encounter each element in the netlist of the network, we will add the specialized stamps corresponding to that element into the sparse system matrix. In this example (see Fig. \ref{fig:equivalent_circuit_w_companion}), there are three ideal voltage sources, six resistance and six inductor models. Therefore, we will cover the stamps for these elements, which will be sufficient to construct the linearized nodal equations for each time-step $(Y_tV_t-J_t=0)$ for the network in Fig. \ref{fig:equivalent_circuit_w_companion}. By recursively solving these equations over time from the initial time-step $t_0$ to the final time step $t_{final}$, we can obtain the time-domain response of this network. Next, we will cover the stamps for these elements. 
\vskip 1em
\noindent Note: We will refer to the sparse linear matrix for the system as $Y$ from here on. Each time step will have a different system matrix notated by $Y_t$ or sometimes simply by $Y$. Further, even for each time step, the matrix can be expressed as the sum of two separate matrices: $Y_t = Y_t^{lin} + Y_t^{nlin}$. Terms in $Y_t^{nlin}$ change for each iteration of NR, whereas the $Y_t^{lin}$, remain fixed. Because this circuit is linear, we do not have $Y_t^{nlin}$.

\subsubsection{Stamp for resistors}

\begin{figure}[htp]
    \centering
    \includegraphics[width=16cm]{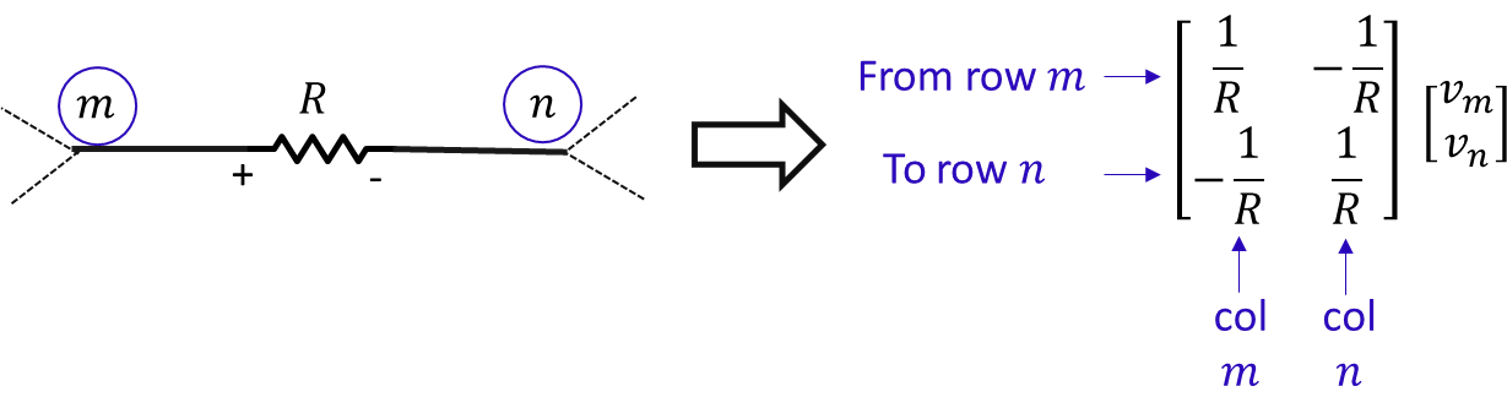}
    \caption{Resistance stamps in the nodal matrix. Redrawn from CMU, ECE, 18-762 Notes \cite{circuit_simulation_pileggi}.}
    \label{fig:resistor_stamp}
\end{figure}

The stamps for a resistance $R$ between node $m$ and $n$ in the nodal matrix $Y$ are shown in Fig. \ref{fig:resistor_stamp}. The general approach is that when you encounter resistance in an input file, add the terms defined in Fig. \ref{fig:resistor_stamp} to the sparse linear matrix $Y$ in the index locations: $(m, m), (m, n), (n, m),$ and $(n, n)$. Here node $m$ is the from node of the resistor, and $n$ is the to node of the resistor. 

%\begin{wrapfigure}{L}{0.5\textwidth}
%\end{wrapfigure}

\subsubsection{Stamps for an independent current source}

\begin{figure}[htp]
    \centering
    \includegraphics[width=10cm]{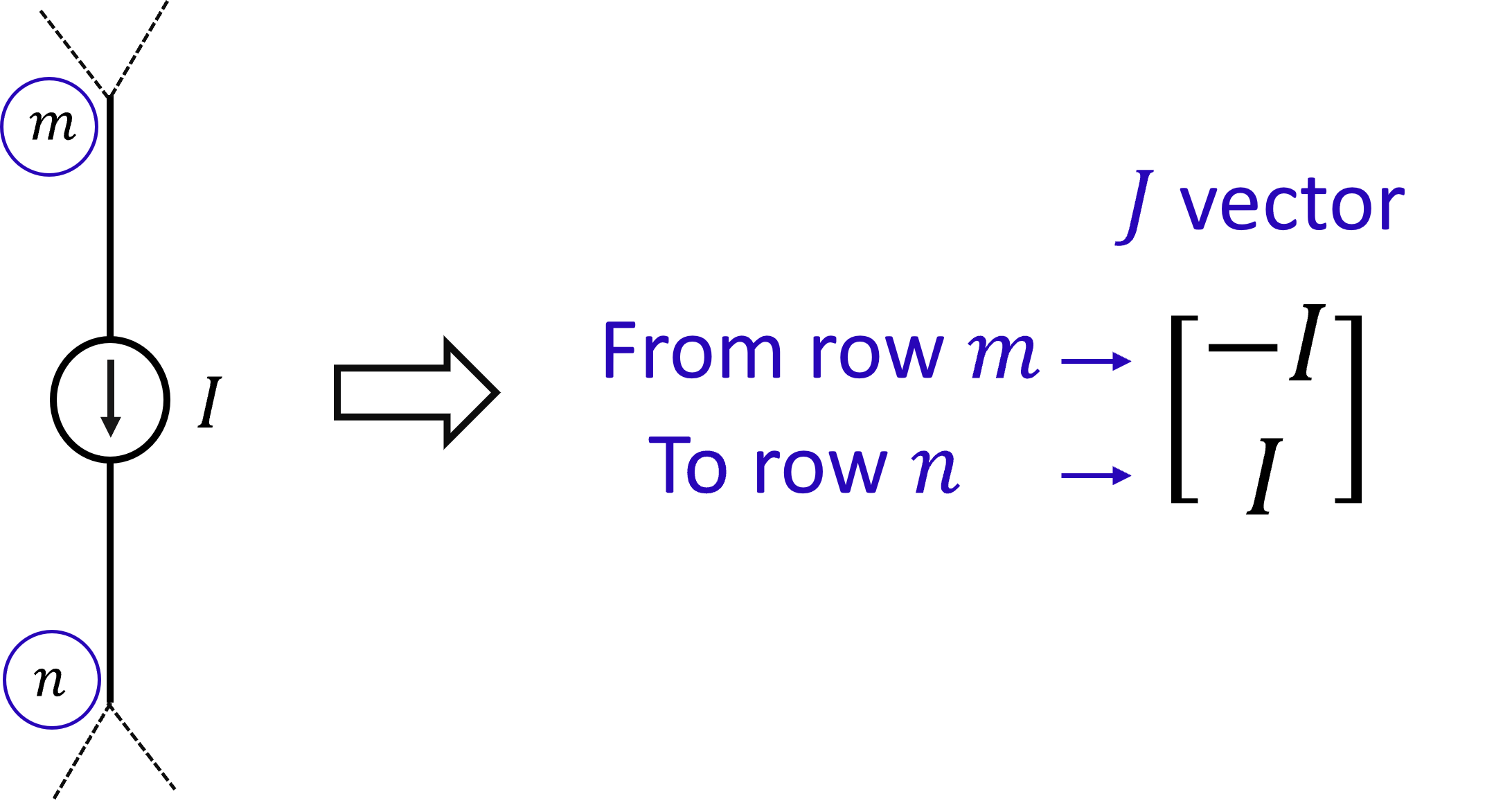}
    \caption{Independent current source stamps in J vector. Redrawn from Pileggi, CMU 18-762 Notes \cite{circuit_simulation_pileggi}.}
    \label{fig:current_stamp}
\end{figure}

An independent current source stamp $I$ between node $m$ and $n$ will add no additional stamps to matrix $Y$ but instead will add terms to column vector $J$ as shown in Fig. \ref{fig:current_stamp}. The column vector $J$ only consists of constant terms that are not a function of any state in the network. The negative sign in the $J$ vector is due to moving the constant number to the RHS of the equation. However, keep in mind the direction of the current flow from the independent current source for the signage.

%\begin{wrapfigure}{L}{0.5\textwidth}
%\end{wrapfigure}

\subsubsection{Stamps for inductors}

Inductor consists of time-derivative terms. Therefore, as a first step,
each time we encounter an inductor in the netlist between node $m$ and node $n$, we will have to stamp the corresponding companion circuit. Remember doing so will replace the time-derivative terms with algebraic expressions. The companion circuit consists of a current source and a resistor in parallel between nodes $m$ and $n$ as seen in Fig. \ref{fig:inductor_companion_circuit}. As we have already covered stamps for resistance and current source, follow the steps in the previous subsections to stamp those components. The key thing to note is that stamp values for inductors in system matrix $Y$ will change for each time step and must be updated accordingly. This is because the values of the independent current source in the companion circuit will be a function of currents and voltages from the last time tick $t_n$.

\subsubsection{Stamps for independent voltage source}

\begin{figure}[htp]
    \centering
    \includegraphics[width=14cm]{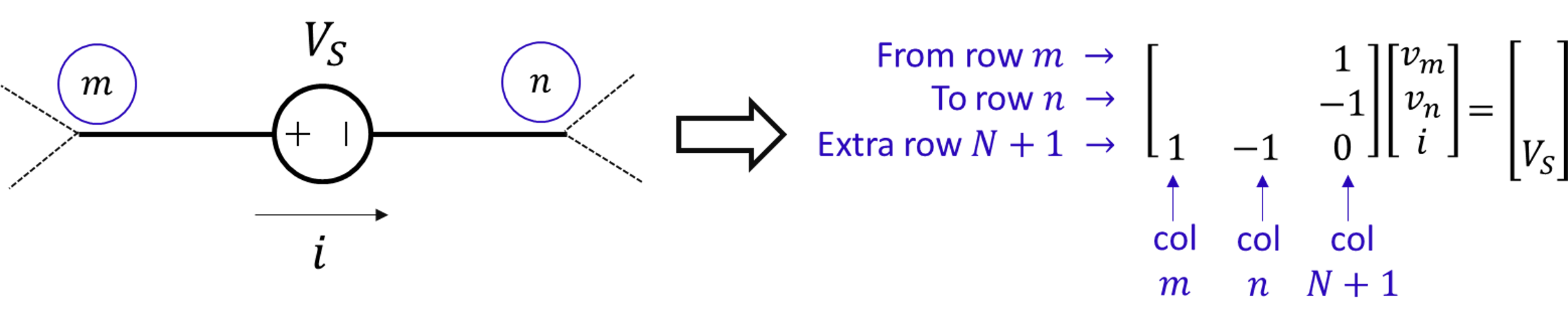}
    \caption{Independent voltage source stamps. Redrawn from Pileggi, CMU- 18-762 Notes \cite{circuit_simulation_pileggi}.}
    \label{fig:voltage_stamp}
\end{figure}

Next, we derive the stamps for a voltage source. To include an independent voltage source between nodes $m$ and $n$, nodal equations are augmented to include an additional row. With an extra row comes an additional variable representing the magnitude of current flow through the voltage source. The additional row (for each voltage source) constrains the potential difference between node $m$ and node $n$:

\begin{align}
    V_m - V_n = V_{S}
\end{align}

\noindent, and the additional current is notated by $i$. The stamps in the nodal matrix $(Y)$ and source vector $(J)$ for independent voltage source are shown in Fig. \ref{fig:voltage_stamp}

%\begin{wrapfigure}{L}{0.5\textwidth}
%\end{wrapfigure}

Now that we have learned how to develop stamps for the resistor, inductor, current source, and voltage source, we can construct the nodal matrix $(Y)$ and source vector $(J)$ for the 
simple power system network in Fig. \ref{fig:ideal_voltage_network_RL_node} for each time $t_{n+1}$. To obtain the time-domain response of the network, the set of linearized equations is recursively solved with a sparse matrix solver (using LU factorization). Note that to obtain the network's response at time $t=0^+$, we require the network states at time $t_n = 0$. For this particular network, these are needed to initiate the terms in inductor stamps dependent on the prior time step. Not just that, it also helps dictate the system state at which the time-domain response begins.

\subsubsection{Initialization}

We can initialize the linear circuit through AC or DC analysis. In DC analysis, we replace the sinusoidal voltage sources with DC voltage sources. In AC analysis, we convert the inductors to an impedance at a single frequency.

Here we will describe the DC analysis approach to initialization. Let us assume the network is in the DC state between $t$ = $-\infty$ and $t=0$. At $t=0$, it switches to the AC sinusoidal source, and let us assume the switching is smooth (i.e., DC voltage and AC voltage at time $t=0$ are identical). To know the initial states of the circuit at $t=0$, we perform DC analysis, wherein the following condition holds true for memory elements (inductors and capacitors):

\begin{align}
     L\frac{di}{dt} = 0 \quad @DC \label{L_DC}\\
     C\frac{dv}{dt} = 0 \quad @DC \label{C_DC}
\end{align}

\begin{figure}[H]
    \centering
    \includegraphics[width=14cm]{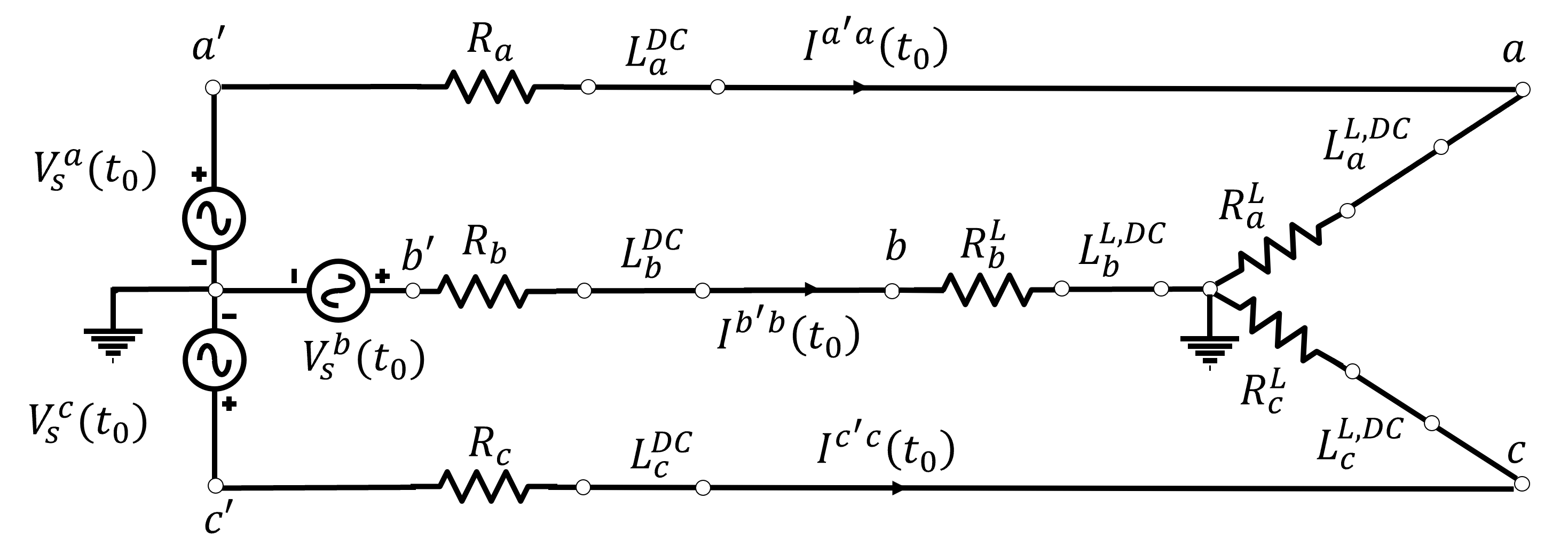}
    \caption{Corresponding equivalent circuit during DC state. Note that the sinusoidal voltage sources are set to a fixed DC value in the case of DC analysis.}
    \label{fig:DC_RL_circuit}
\end{figure}

\noindent From these conditions in \eqref{L_DC} and \eqref{C_DC}, we can infer that during the DC state, the inductor $L^{DC}$ is short-circuited $(V_L = 0)$ and the capacitor $C^{DC}$ is open-circuited $(I_C = 0)$. For this example, the DC equivalent of the network in Fig. \ref{fig:equivalent_circuit_w_companion} is shown via a network in Fig. \ref{fig:DC_RL_circuit}. With access to DC equivalent circuit, we stamp the elements in the circuit into the system matrix $Y_{dc}$ and vector $J_{dc}$ and solve them using a sparse linear solver. The solution is used to define the system state at $t=0$.

%\begin{wrapfigure}{L}{0.5\textwidth}
%\end{wrapfigure}

\subsubsection{Time-domain solution} \label{sec:linear_analysis}

\begin{figure}[H]
    \centering
    \includegraphics[width=14cm]{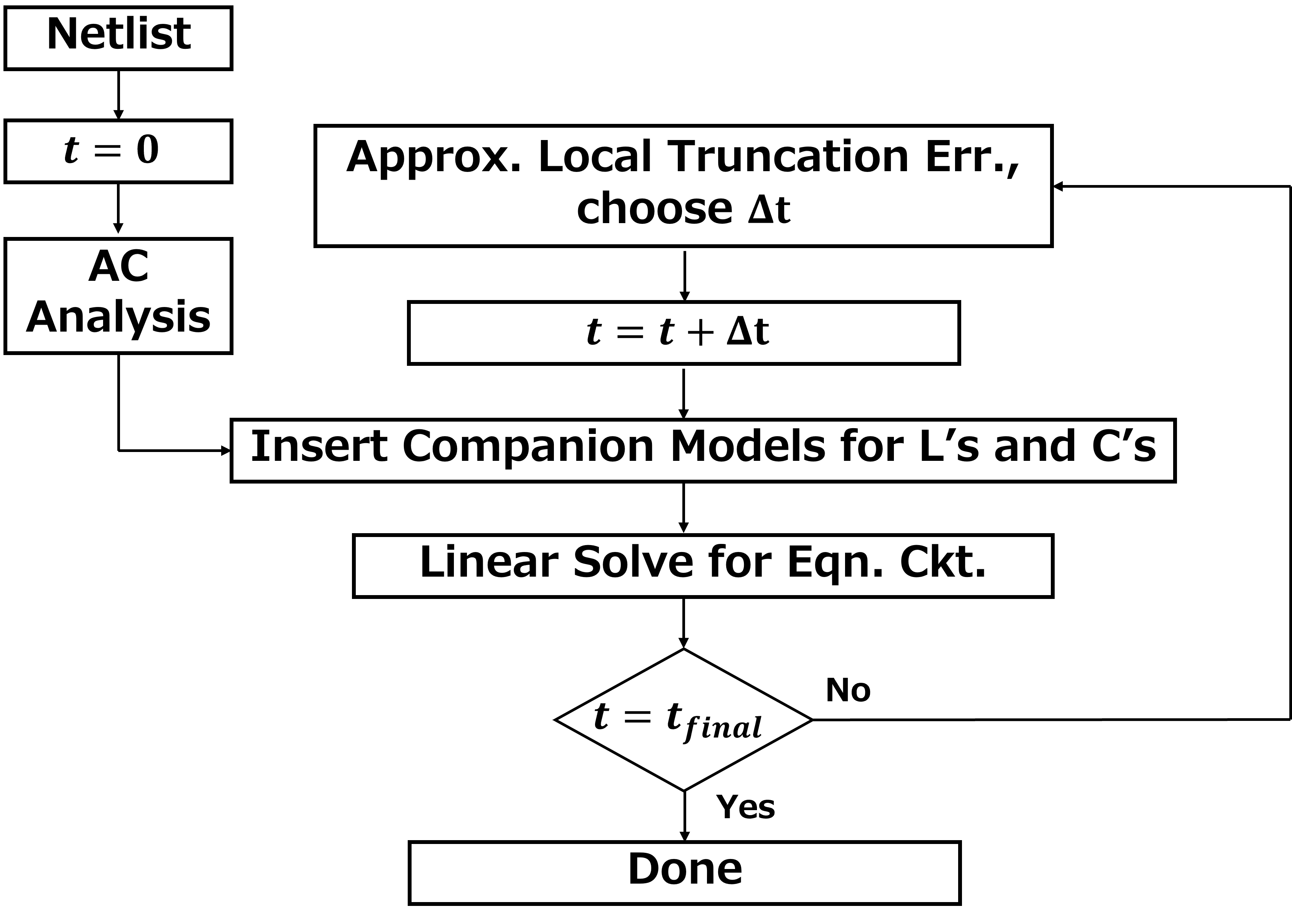}
    \caption{Running EMT simulation for linear networks.}
    \label{fig:linear_analysis_flowchart}
\end{figure}

 With access to the initial state at $t=0$, we recursively solve a system matrix (moving forward by $\Delta t$ in each recursion) following steps described in Fig. \ref{fig:linear_analysis_flowchart} to obtain the time-domain response from $t=0$ to $t=t_{final}$. We can dynamically adjust $\Delta t$ at each step based on the trade-off between simulation run-time and local truncation error (LTE). LTE is calculated by approximating the higher-order terms of the numerical integration. More details can be found in \cite{circuit_simulation_pileggi}. Note that dynamically adjusting LTE is not always necessary as for a purely linear system changing $\Delta t$ will require re-generating the $\Delta t$ dependent companion model terms at each iteration at an additional computational cost. Practically speaking, a happy medium can be obtained by periodically checking the LTE throughout the simulation and adjusting $\Delta t$ if it is found high.

One facet that this example does not cover is how to handle nonlinear terms. We will learn how to include nonlinear terms in the following example with IM as an electric load. We will also learn to use an alternative mathematical form (other than MNA) to describe the system's behavior. In addition, we will also learn how to incorporate non-electrical physics (i.e., IM's mechanical physics), like rotor speed, with the circuit-theoretic paradigm, into the mathematical formulation.

\subsection{Induction Motor Load}

\begin{figure}[H]
    \centering
    \includegraphics[width=14cm]{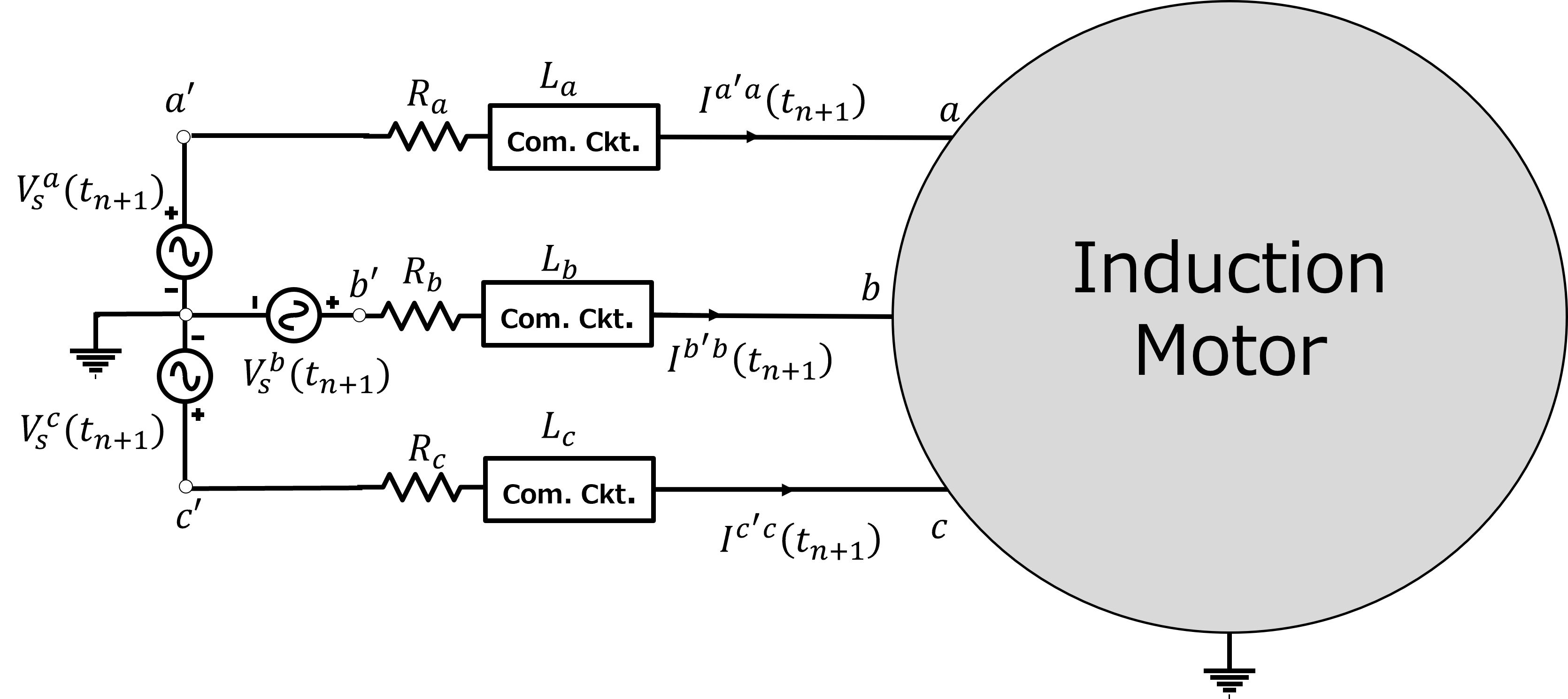}
    \caption{2-bus network with induction motor load.}
    \label{fig:IM_load}
\end{figure}

In the second example, we will learn how to obtain the time-domain response for a non-linear grid component: the induction motor (IM). To construct a toy power grid network with IM, we will replace the RL load in the first example with an IM load (see Fig. \ref{fig:IM_load}).

%\begin{wrapfigure}{L}{0.5\textwidth}
%\end{wrapfigure}

Analyzing IM in the \textit{abc} frame can be difficult. The flux generated by the three-phase IM in \textit{abc} frame has time-varying coefficients in its voltage terms due to the sinusoidal nature of the mutual inductance. This makes the analysis of three-phase IM cumbersome in the \textit{abc} reference frame. However, this undesirable feature can be eliminated by using \textit{dq} transformation. \textit{dq} transformation is a linear transformation. It can be performed by choosing one of the three reference frames: i) synchronous reference frame, ii) stationary reference frame, and iii) rotating reference frame. For more details on dq transformation, see \cite{lee_dq}.

For derivations in this document, we will use the stationary reference frame and a power invariant dq transformation. Doing so allows us to set $\theta = 0$ and simplify further calculations by having a time-invariant voltage source across the IM. In the stationary frame,  \textit{abc} variables (both currents and voltages) are converted to \textit{dq} variables by following power invariant dq matrix transformation:

\begin{align}
     [F_{0dq}] = [P_{\theta}] [F_{abc}]
\end{align}

\noindent where:

\begin{align}
    [P_{\theta}] = \sqrt{\frac{2}{3}} \begin{bmatrix}
    \sqrt{0.5} & \sqrt{0.5} & \sqrt{0.5}\\
    cos(\theta) & cos(\theta - \lambda) & cos(\theta + \lambda)\\
    sin(\theta) & sin(\theta - \lambda) & sin(\theta + \lambda)
    \end{bmatrix}, \theta = 0
\end{align}

\noindent $\lambda$ is the phase difference between phases $abc$ during balanced operation:

\begin{align}
    \lambda = \frac{2\pi}{3} rad
\end{align}

\begin{figure}[H]
    \centering
    \includegraphics[width=15cm]{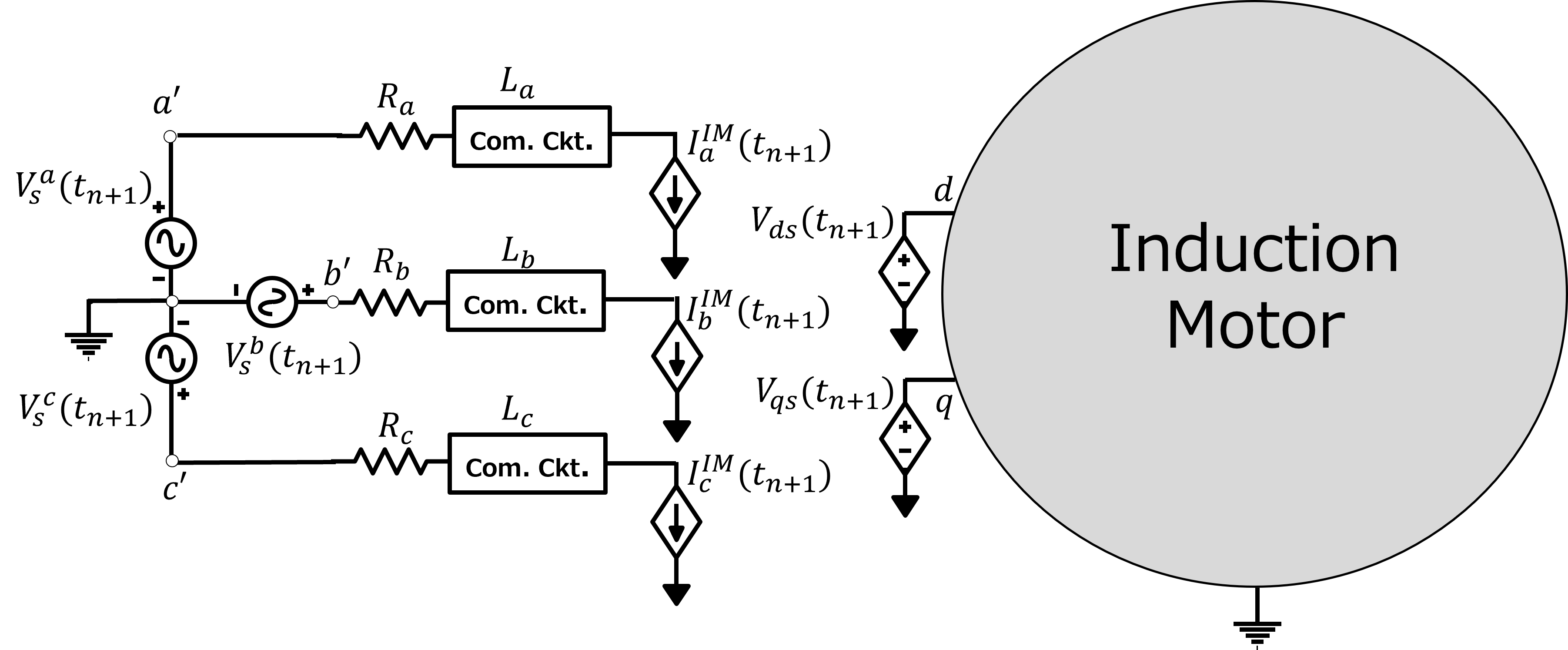}
    \caption{2-bus network with induction motor load after dq transformation.}
    \label{fig:IM_dq_load}
\end{figure}

\noindent After applying \textit{dq} transformation to IM and decoupling non-IM \textit{abc} components (see left of Fig. \ref{fig:IM_dq_load}) from \textit{dq} components using controlled current and voltage sources, the network model in Fig. \ref{fig:IM_load} is modified to one in Fig. \ref{fig:IM_dq_load}.

Note that controlled current and voltage sources in Fig. \ref{fig:IM_dq_load} encapsulate the math behind \textit{dq} and inverse \textit{dq} transformation for the variables. \textit{dq} transformation is applied to capture \textit{abc} network voltages and reflect those for IM sub-circuit in \textit{dq} frame. Inverse \textit{dq} transformation is used to convert the currents consumed by the IM in \textit{dq} frame into \textit{abc} frame ($I_a^{IM}, I_b^{IM}, I_c^{IM})$ such that they can be fed into rest of the network.

%\begin{wrapfigure}{L}{0.5\textwidth}
%\end{wrapfigure}

The transformed Fig. \ref{fig:IM_dq_load} has three components: i) the infinite bus represented by a set of wye-connected independent voltage sources, ii) the transmission line represented by a series RL circuit, and iii) an IM. 
We have already discussed the construction of stamps and companion circuits for infinite bus and transmission line elements (reiterating that in this tutorial the transmission line model is greatly simplified, see \cite{dommel1996emtp} for detailed models like pi model and Bergeron model). 
Now, we will learn how to represent the IM physics with an equivalent circuit. 
We will also learn how to replace the time derivative terms in the IM equivalent circuit with an analogous companion circuit.

The generalized equations that represent the electrical component of IM physics, independent of the reference frame choice in dq transformation, are expressed via the following KVL equations \cite{lee_dq}:

\begin{align}
    V_{ds}=R_sI_{ds}+p\psi_{ds}+\psi_{qs}p\theta \label{eq:IM_elec_first} \\
    V_{qs}=R_sI_{qs}+p\psi_{qs}-\psi_{ds}p\theta\\
    V_{dr}=R_rI_{dr}+p\psi_{dr}+\psi_{qr}p\beta\\ 
    V_{qr}=R_rI_{qr}+p\psi_{qr}-\psi_{dr}p\beta
\end{align}

\noindent where $p$ is the differential operator and,

\begin{align}
    \psi_{ds}={(L}_{ls}+L_m)I_{ds}+L_mI_{dr}\\
    \psi_{dr}={(L}_{lr}+L_m)I_{dr}+L_mI_{ds}\\
    \psi_{qs}={(L}_{ls}+L_m)I_{qs}+L_mI_{qr}\\
    \psi_{qr}={(L}_{lr}+L_m)I_{qr}+L_mI_{qs} \label{eq:IM_elec_last}
\end{align}

In the expressions above, $L_{ls}$ and $L_{lr}$ represent the leakage inductance of the stator circuit and rotor circuit, respectively. $L_m$ is the mutual inductance between the rotor and stator circuits. $R_s$ and $R_r$ are the stator and rotor resistance, respectively. $\psi$ represents the flux component across each electrical sub-circuit. The nonlinearity in the electrical part of the IM is due to the speed voltage terms ($\psi_{qr}\omega_r, \psi_{dr}\omega_r$).

In addition to the IM electrical part equations in \eqref{eq:IM_elec_first}-\eqref{eq:IM_elec_last}, the mechanical IM part is represented by a single differential equation, the swing equation \eqref{eq:IM_mech}:

\begin{align}
    p\omega_r=\frac{(T_E-T_L-D\omega_r)}{J}	\label{eq:IM_mech}
\end{align}

\noindent where, electrical torque $T_E$ is further described by interactions between the IM currents in the electrical sub-circuit, which also introduces nonlinearities in the IM model:

\begin{align}
    T_E=\frac{3}{4}L_mN_p(I_{dr}I_{qs}-I_{qr}I_{ds}) \label{eq:T_e}
\end{align}

\noindent $T_E$, the electrical torque, is given in $N.m$ and $J$, the motor net inertia, in $kg.m^2.$ $N_p$  is the number of poles in the IM. 
The load torque $T_L$ is generally represented as a polynomial function of rotor speed.

We can further simplify the electrical components' expressions because we use the stationary reference frame for dq transformation. In the stationary reference, $\theta=0$, $\beta=-\theta_r$ and $p\beta=-\omega_r$. Furthermore, $p\theta = \omega_s = 0$. With these simplifications and reducing the expressions in \eqref{eq:IM_elec_first}-\eqref{eq:IM_elec_last}, the following form for the IM's electrical part, in the stationary frame, can be obtained:

\begin{align}
    V_{ds} - R_sI_{ds} - L_s pI_{ds} - L_m p I_{dr} = 0 \label{eq:IM_i} \\
    V_{qs} - R_sI_{qs} - L_s pI_{qs}  - L_m p I_{qr} = 0
    \label{eq:IM_ii} \\
    V_{dr} - R_rI_{dr} - L_r pI_{dr} - L_m p I_{ds} + 
    \omega_r L_r I_{qs} + \omega_r L_m I_{qr} = 0
    \label{eq:IM_iii}\\
    V_{qr} - R_r I_{qr} - L_r pI_{qr} - L_m p I_{qs} - 
    \omega_r L_r I_{dr} - \omega_r L_m I_{ds} = 0
    \label{eq:IM_iv}
\end{align}

\noindent where,

\begin{align}
    L_s = Lls + L_m\\
    L_r = Llr + L_m\\
    V_{dr} = 0\\
    V_{qr} = 0
\end{align}

\noindent Note that the rotor windings are shorted in most IM designs, and hence voltages $V_{dr}$ and $V_{qr}$ are set to 0.

\begin{figure}[H]
    \centering
    \includegraphics[width=14cm]{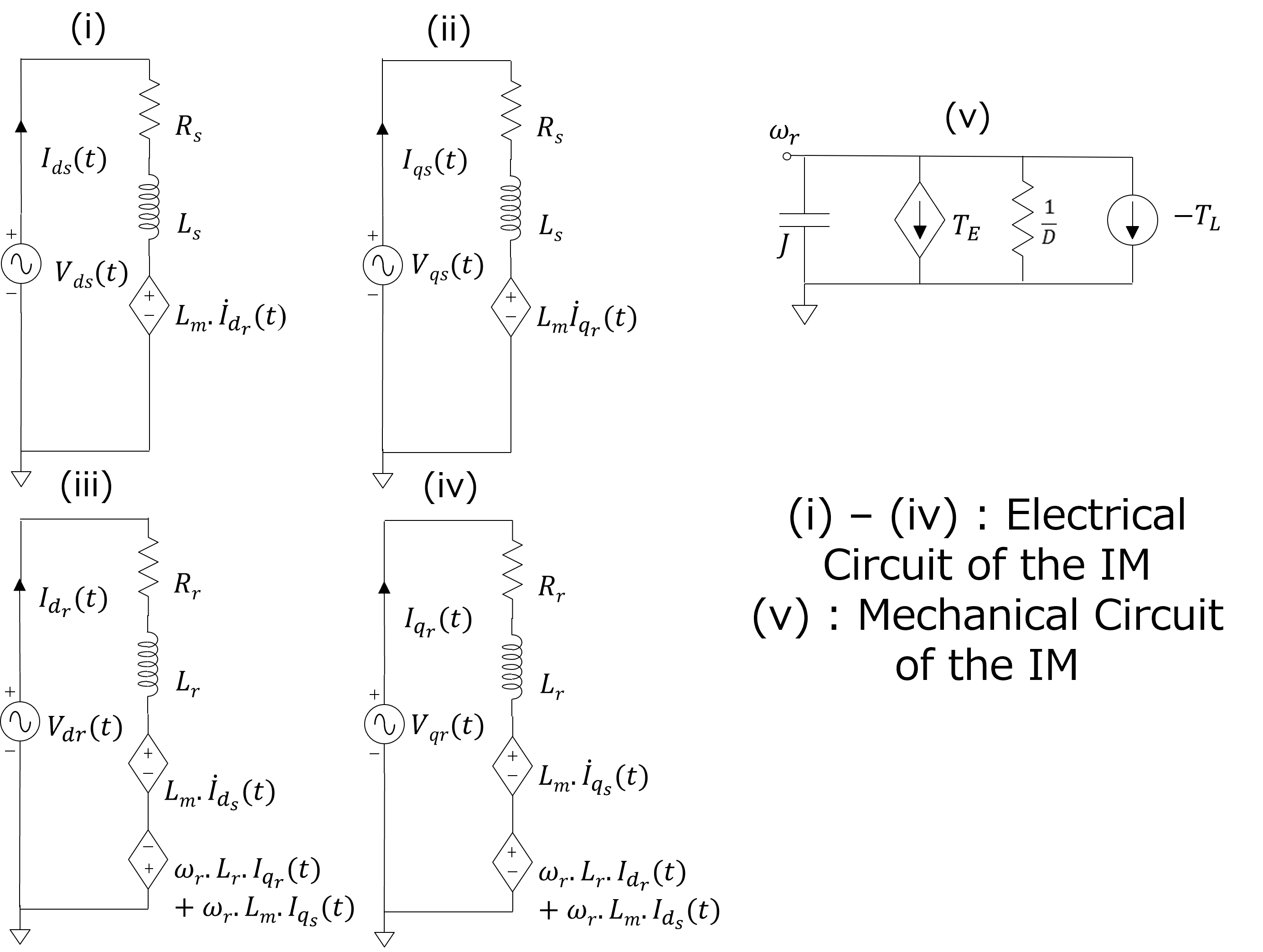}
    \caption{Equivalent circuit for IM in the stationary frame.}
    \label{fig:IM_equi_circuit}
\end{figure}

Now to construct the equivalent circuit for the IM model, we map the four KVL expressions in \eqref{eq:IM_i}-\eqref{eq:IM_iv} and one KCL expression in \eqref{eq:IM_mech} to Fig. \ref{fig:IM_equi_circuit}.

%\begin{wrapfigure}{L}{0.5\textwidth}
%\end{wrapfigure}

\begin{figure}[H]
    \centering
    \includegraphics[width=14cm]{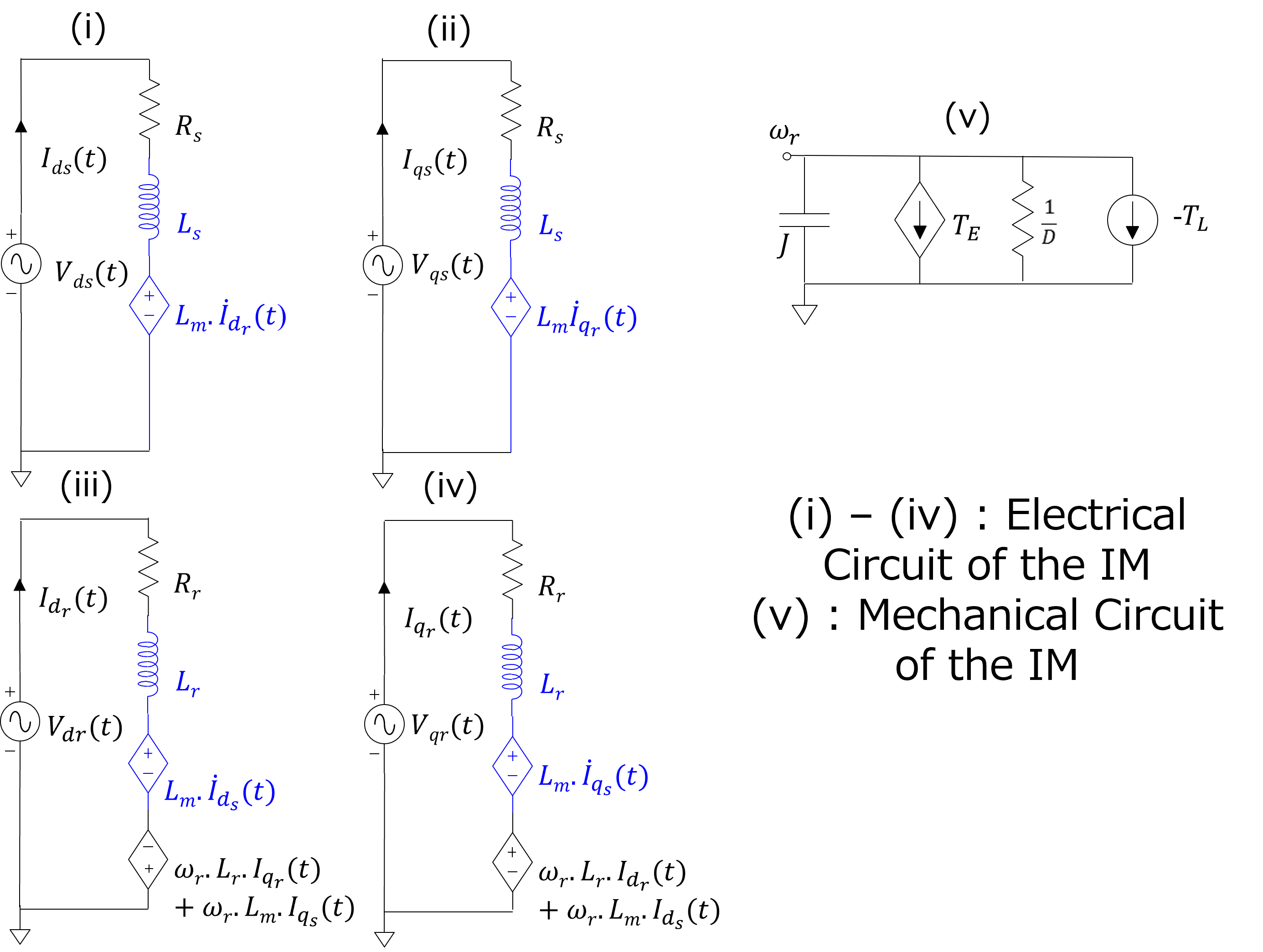}
    \caption{Illustration of combined self and mutual inductances in IM.}
    \label{fig:IM_equi_circuit_mutual}
\end{figure}

\begin{figure}[H]
    \centering
    \includegraphics[width=10cm]{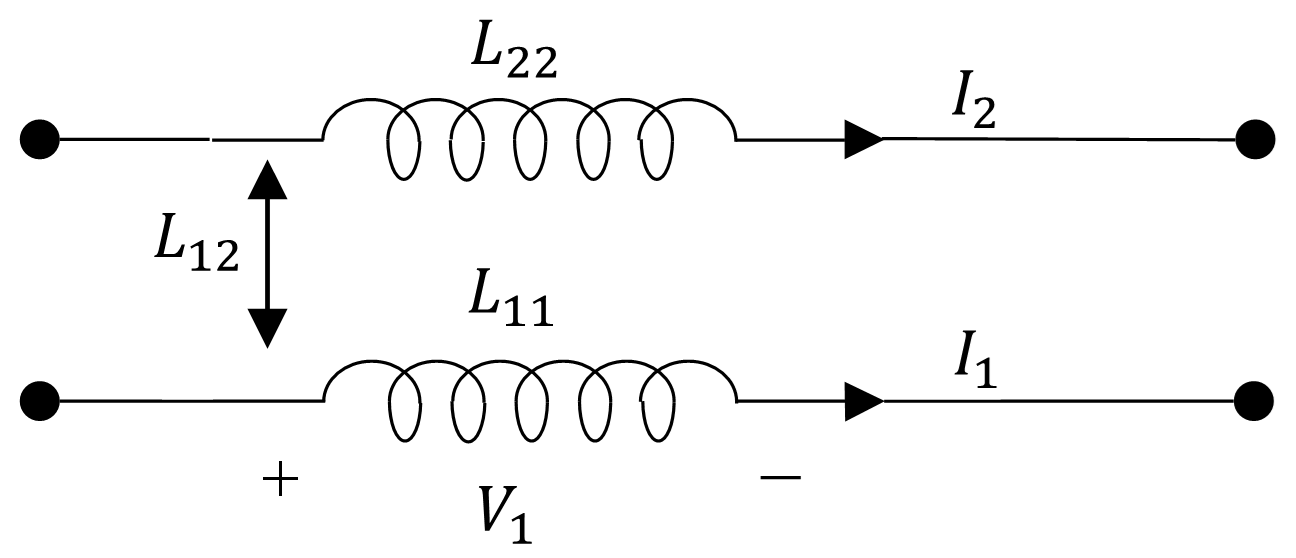}
    \caption{Two coil example.}
    \label{fig:two_coil}
\end{figure}

One can observe in Fig. \ref{fig:IM_equi_circuit} that aside from self-inductance ($L_s$ and $L_r$), the IM equivalent circuit also includes time-derivative terms (given by $p$ or $\dot I / \dot \omega_r$) for mutual inductance $(L_m)$ and rotor speed $(\omega_r)$. We have previously learned how to approximate time-derivative terms for self-inductance by constructing and stamping the corresponding companion circuits. Here, we will derive the companion circuits for combined self and mutual inductance elements (see blue elements in Fig. \ref{fig:IM_equi_circuit_mutual}). We will use a simple two-coil example in Fig. \ref{fig:two_coil} to develop the companion circuit for combined self- and mutual inductance. Later in the section, we will learn how to construct the companion circuit for time-derivative terms corresponding to rotor speed. We will finish by discussing how the nonlinear elements in the IM equivalent circuit are handled and subsequently added to the system matrix.

%\begin{wrapfigure}{L}{0.5\textwidth}
%\end{wrapfigure}

We construct the companion circuit for combined self and mutual inductances to approximate its time-derivative behavior by a set of algebraic equations. This way, we can approximately solve a set of ODEs parameterized by self- and mutual- inductances by recursively solving a set of algebraic equations over time. We begin the derivation by observing a two-coil example in Fig. \ref{fig:two_coil}, which has two pairs of self- and mutual-inductance (self: {$L_{11}$ or $L_{22}$}, mutual: {$L_{12}$}). We focus on voltage $V_1$ expression in the first coil. It can be represented as a sum of the voltage across the self and mutual inductances:

\begin{align}
    V_1(t) = L_{11}\frac{{dI}_1}{dt}+L_{12}\frac{{dI}_2}{dt} \label{eq:2coil}
\end{align}

\noindent The voltage expression in \eqref{eq:2coil} is an ODE that includes time-derivative terms. Therefore, we apply the trapezoidal integral rule to approximate the ODE's solution. With the trapezoidal integral rule applied, the voltage across self- and mutual- inductance \eqref{eq:2coil} can be represented by the sum of the following difference equations, which are purely algebraic terms:

\begin{align}
    V_{11}(t_{n+1}) = \frac{2L_1}{\Delta t}\left(I_1(t_{n+1}) - I_1(t_n)\right) - V_{11}(t_n)\\
    V_{12}(t_{n+1}) = \frac{2L_{12}}{\Delta t}\left(I_2(t_{n+1}) - I_2(t_n)\right) - V_{12}(t_n)
\end{align}

\noindent Therefore the total voltage induced in coil 1 $(V_1(t) = V_{11}(t) + V_{12}(t))$ is given by:

\begin{align}
    V_{1}(t_{n+1}) = \frac{2L_1}{\Delta t}\left(I_1(t_{n+1}) - I_1(t_n)\right) - V_{11}(t_n) + \frac{2L_{12}}{\Delta t}\left(I_2(t_{n+1}) - I_2(t_n)\right) - V_{12}(t_n)
\end{align}

\noindent and after re-arranging the terms:

\begin{align}
    V_{1}(t_{n+1}) = \color{blue} \frac{2L_1}{\Delta t}I_1(t_{n+1}) +
    \color{red}
    \frac{2L_{12}}{\Delta t}I_2(t_{n+1})  
    \color{olive}
    - \left( \frac{2L_1}{\Delta t} I_1(t_n) + \frac{2L_{12}}{\Delta t} I_2(t_n) + V_{12}(t_n) + V_{11}(t_n)\right) \label{eq:two_coil_volt}
\end{align}

\begin{figure}[H]
    \centering
    \includegraphics[width=10cm]{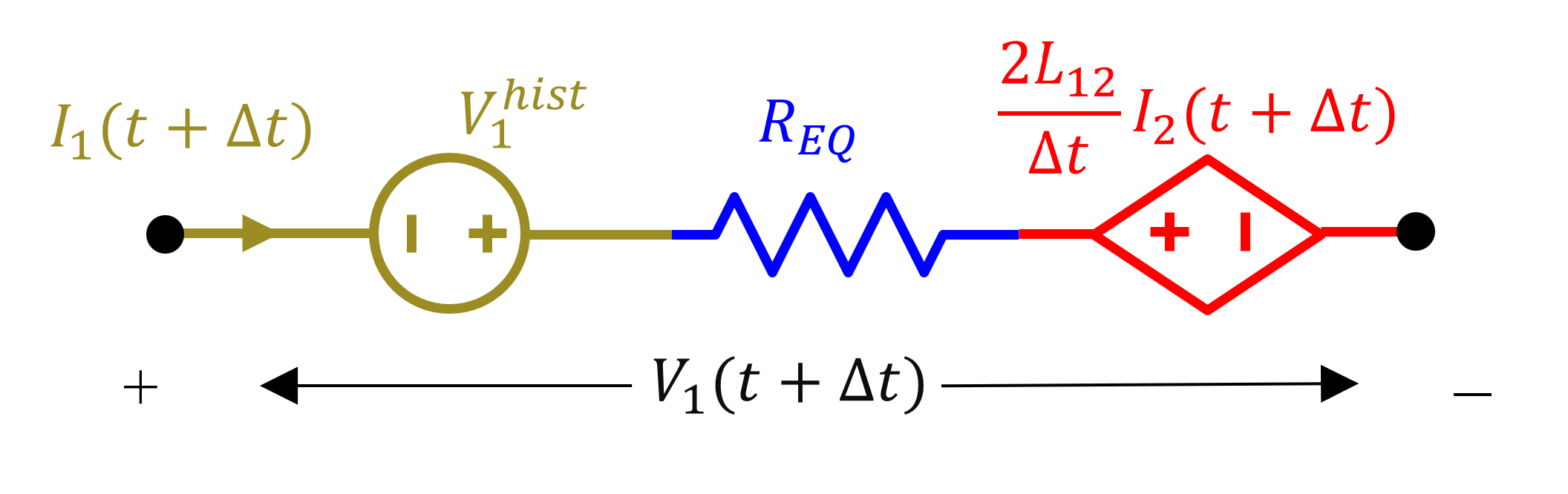}
    \caption{Two coil example companion circuit.}
    \label{fig:two_coil_companion}
\end{figure}

The voltage at time $t_{n+1}$ across the coil 1 is approximated using \eqref{eq:two_coil_volt} and is represented as an equivalent circuit in Fig. \ref{fig:two_coil_companion}. The \textcolor{olive}{third term} is only dependent on the historical values of the variables and can be represented by an \textcolor{olive}{independent voltage source $V_{1}^{hist}$}. Similarly, a \textcolor{blue}{resistance $R_{EQ}$} represents \textcolor{blue}{first term} as it maps the linear relationship between the voltage and current across the same branch and a \textcolor{red}{current controlled voltage source} represents the \textcolor{red}{second term}, as it maps the voltage contribution due to current in the second coil. 

%\begin{wrapfigure}{L}{0.5\textwidth}
%\end{wrapfigure}

By replacing the time-derivative terms for self- and mutual-inductance with the corresponding companion circuits in Fig. \ref{fig:two_coil_companion}, we can solve the set of ODEs corresponding to the electrical part of IM, recursively over time. But before, we must learn how to address the time-derivative term in the mechanical part of the IM model. The variables in the mechanical part impact the electrical part and vice-versa. Therefore we need to solve the mechanical part concurrently with the electrical portion of IM. The approach for solving the mechanical part is trivial. On close observation, one can see the mechanical circuit of the IM is analogous to a parallel RC electrical circuit with current sources, where inertia $J$ is analogous to capacitance $C$, $T_L$ is analogous to a constant current source $I$, $D$ is analogous to a conductance $G$, $T_E$ is analogous to a current controlled current source (see \eqref{eq:T_e}), and $\omega_r$ is analogous to the voltage at the mechanical circuit. The only time-derivative term in the mechanical circuit is the partial derivative of rotor speed times inertia $(J\frac{d\omega_r}{dt})$, which is analogous to a current through a capacitor. Therefore, we will learn how to construct a capacitor's companion circuit to replace the time-derivative term in the mechanical part of IM. By replacing the time-derivative term, we can represent the physics of IM's mechanical part with a single nodal equation that can be solved recursively over time (assuming we have dealt with nonlinearities). 

To construct the companion model for a capacitor, we will use trapezoidal integration approximation following the same approach we used for inductors. 
Applying the trapezoidal integration approximation, the expression for current/voltage through/across a capacitor at time $t_{n+1}$ given values at time $t_n$ can be expressed as follows:

\addtolength{\jot}{1em}
\begin{align} 
    C\frac{dv}{dt} = i\\
    dv = \frac{idt}{C}\\
    \int_{t_n}^{t_{n+1}}dv =  \int_{t_n}^{t_{n+1}}\frac{idt}{C}\\
    \left(v(t_{n+1}) - v(t_n) \right) = \left(i(t_{n+1}) + i(t_n)\right)\frac{\Delta t}{2C}\\
    v(t_{n+1}) = v(t_n) + i(t_n)\frac{\Delta t}{2C} + i(t_{n+1})\frac{\Delta t}{2C} \label{eq:cap}
\end{align}

\begin{figure}[htp]
    \centering
    \includegraphics[width=14cm]{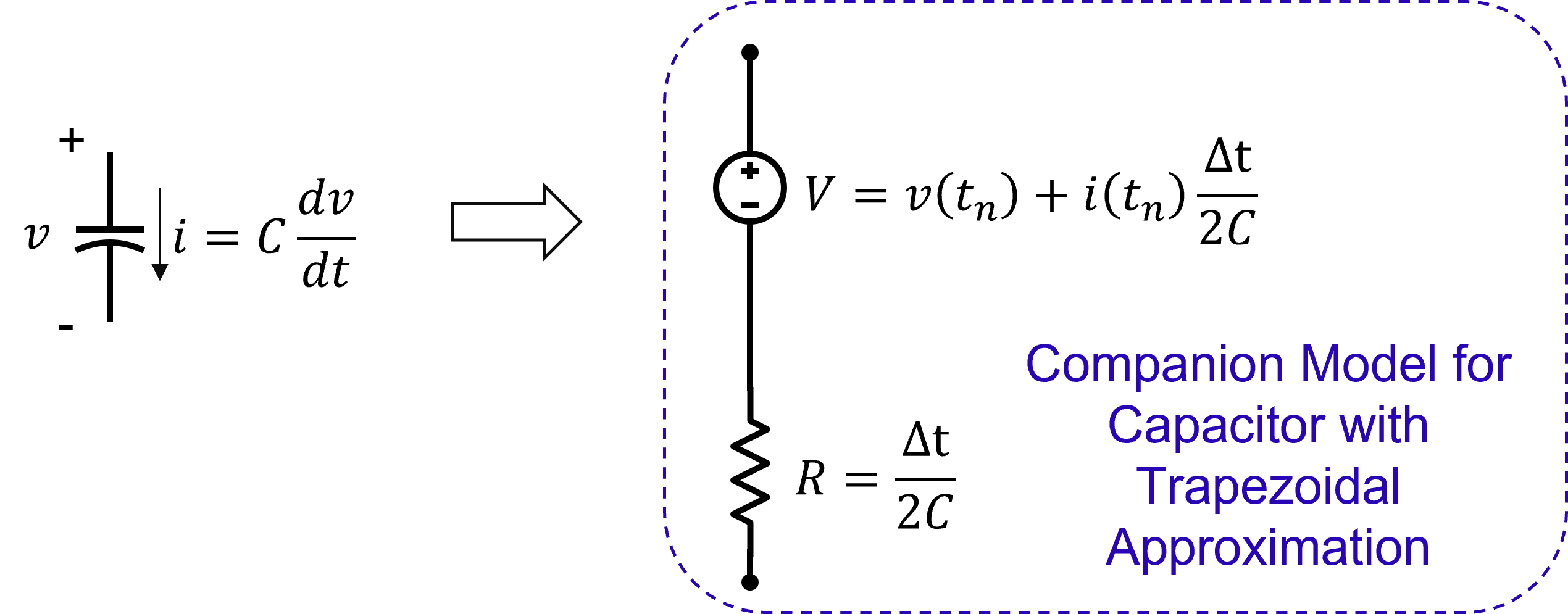}
    \caption{Capacitor Trapezoidal Companion Circuit. Redrawn from Pileggi, 18-762 Notes \cite{circuit_simulation_pileggi}.}
    \label{fig:capacitor_companion_circuit}
\end{figure}

\noindent The equation in \eqref{eq:cap} has three terms. The last term $\frac{\Delta t}{2C}$ describes the relationship between current and voltage at time $t_{n+1}$ and is a conductance $G$. The first two terms are constants that are known from prior time-point $t_n$ and therefore are represented via a constant voltage source $V$. Aggregating these elements in series, we get the equivalent circuit in Fig. \ref{fig:capacitor_companion_circuit}, representing a capacitor's companion model with trapezoidal numerical integration approximation. For representing the speed time-derivative terms in IM, we will use analogous symbols in the companion model: inertia $J$ for capacitor $C$ and rotor speed $\omega_r$ for voltage $v$.

%\begin{wrapfigure}{L}{0.5\textwidth}
%\end{wrapfigure} 

\subsubsection{Handling nonlinear terms}

In the case of IM, we began with a set of models characterized by differential-algebraic equations (DAEs), which had both differential and nonlinear terms. As a first step, we built and used companion circuits to approximate the differential term behavior with algebraic terms. Next, we linearize the nonlinear terms and solve them iteratively to approximate the nonlinear behavior. 

We linearize the nonlinear terms with first-order Taylor's approximation, like in nonlinear power flow analysis. However, instead of linearizing complete nodal or loop equations (e.g., KCL or KVL), we linearize the nonlinear terms corresponding to each component modularly by developing linearized equivalent circuits. Then for each iteration of NR, we iteratively update the stamps corresponding to these components in the system matrix $Y$ and resolve.

Let us write out and understand the general form and application of first-order Taylor approximation. For general nonlinear function $f(x)$ ($x:\mathbb{R}^{N\times1}$), the first order Taylor approximation is given by: 

\begin{align} 
    f^{k+1}(x) \approx f^k (x) + {f^\prime (x)}^T (x^{k+1}-x^k) \label{general_linear}
\end{align}

\noindent With nonlinear $f(x)$ linearized around  $x_k$ from the $k^{th}$ iteration, the next iterate can be obtained via a linear solve. We will use the general theory now to understand the linearization of IM terms.

In the IM model in Fig. \ref{fig:IM_equi_circuit}, for the electrical part, the non-linearity stems from four speed-voltage terms $(\omega_rL_rI_{dr},  \omega_rL_mI_{ds}, \omega_rL_rI_{qr},$ and $\omega_rL_mI_{qs})$ in the $dr$ and $qr$ sub-circuits (iii, and iv in Fig. \ref{fig:IM_equi_circuit}). There are additional nonlinearities in the mechanical sub-circuit (v in Fig. \ref{fig:IM_equi_circuit})

To further explore how to add IM nonlinear terms to the Y matrix, let us consider the nonlinear terms in the KVL expression \eqref{eq:IM_iv} for subcircuit iv) in Fig. \ref{fig:IM_equi_circuit}:

\begin{align} 
f^{NL}_{qr}(\omega_r, I_{ds}, I_{dr}) = -\omega_r L_r I_{dr} - \omega_r L_m I_{ds} \label{eq:nonlinearFqr}
\end{align} 

\noindent \eqref{eq:nonlinearFqr}  is a nonlinear function of the rotor speed $(\omega_r)$, direct-axis rotor current $(I_{dr})$, and direct-axis stator current $(I_{ds})$. The linearized approximation for this expression for the $(k+1)^{th}$ NR iteration is:

\begin{align} 
    f_{qr}^{k+1}(\omega_r, I_{dr}, I_{ds}) = f_{qr}^k (\omega_r, I_{dr}, I_{ds}) + {\left(\frac{\partial f}{\partial \omega_r}\right)}_k  (\omega_r^{k+1}-\omega_r^k) + {\left(\frac{\partial f}{\partial I_{dr}}\right)}_k  (I_{dr}^{k+1}-I_{dr}^k) \\ + {\left(\frac{\partial f}{\partial I_{ds}}\right)}_k  (I_{ds}^{k+1}-I_{ds}^k) \label{eq:linearize_fqr} \notag
\end{align}
\noindent where:
\begin{align} 
    f^k (\omega_r, I_{dr}, I_{ds}) = -\omega^k_r L_r I^k_{dr} - \omega^k_r L_m I^k_{ds} \\
    {\left(\frac{\partial f}{\partial \omega_r}\right)}_k = -L_r I^k_{dr} - L_m I^k_{ds} \\ 
    {\left(\frac{\partial f}{\partial I_{dr}}\right)}_k =  -\omega^k_r L_r \\
    {\left(\frac{\partial f}{\partial I_{ds}}\right)}_k =  -\omega^k_r L_m
\end{align}

Similar to the treatment of nonlinear terms in \eqref{eq:IM_iv}, other nonlinear terms in the IM are also linearized and stored in symbolic form. These terms are added (or stamped) in the system matrix $Y$ following the discussion in next subsection and updated for each iteration of NR.

\subsubsection{Stamping controlled current sources}

In the IM linearized equivalent circuit, we end up with two new circuit elements, a current-controlled voltage source (CCVS) and a current-controlled current source (CCCS). The CCVS originate from linearizing speed current terms in \eqref{eq:IM_i}-\eqref{eq:IM_iv} (see terms in \eqref{eq:linearize_fqr} for $F_{qr}$ KVL constraint). The CCCS terms originate while linearizing the $T_E$ equation. We briefly describe how those terms are added to the system matrix.

\begin{figure}[htp]
    \centering
    \includegraphics[width=10cm]{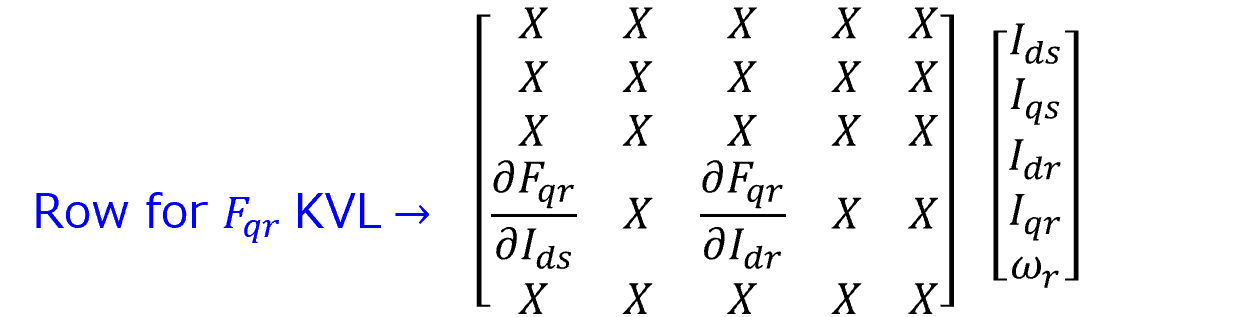}
    \caption{Off-diagonal entries for modeling CCVS in IM's $F_{qr}$ electrical sub-circuit.}
    \label{fig:CCVS}
\end{figure}

As we use KVL to describe the IM's electrical equations, the current-controlled voltage sources only show up as off-diagonal terms in the system matrix. For example, the CCVS stamps for $F_{qr}$ KVL expression are shown in the system matrix in Fig. \ref{fig:CCVS}. These do not include the rotor speed terms.

\begin{figure}[htp]
    \centering
    \includegraphics[width=14cm]{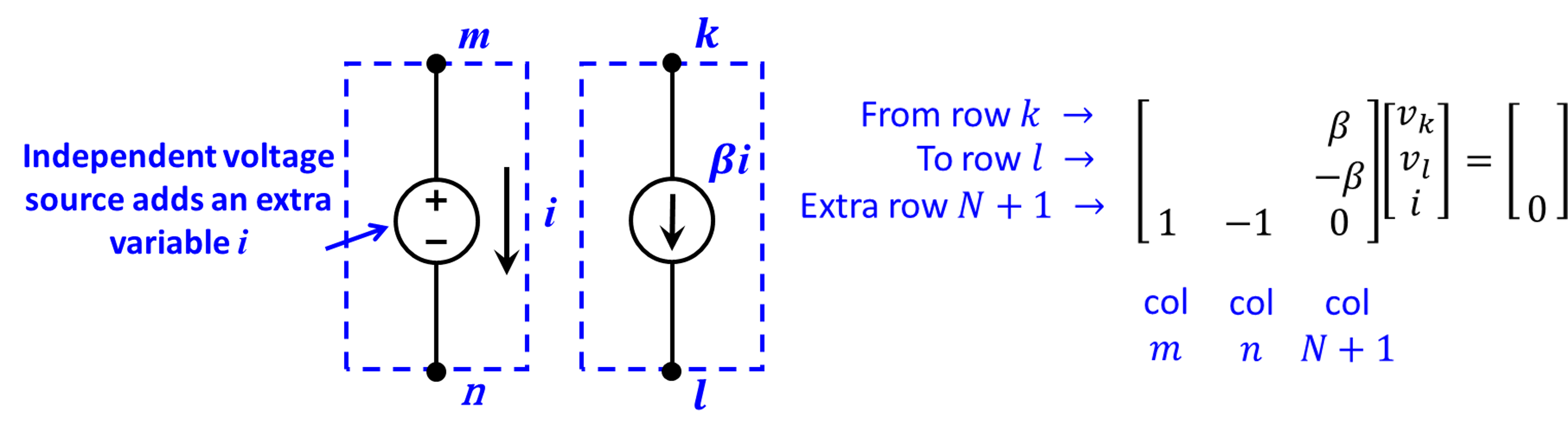}
    \caption{Stamps for CCCS. Reconstructed from Pileggi, Carnegie Mellon ECE 18-762 Notes \cite{circuit_simulation_pileggi}.}
    \label{fig:CCCS}
\end{figure}

Next, we discuss CCCS stamps, which are in the mechanical equation and are stamped following the MNA approach in Example 1. The stamps for a generic CCCS between nodes $k$ and $l$ are shown in Fig. \ref{fig:CCCS}. In general, to measure the independent current on which the controlled source depends, we add a zero-valued voltage source (ammeter) between nodes $m$ and $n$. Note that adding a voltage source results in an addition of a row to the $Y$-matrix, and in return, we get the current value through the source.  Now, as the dependent current source between nodes $k$ and $l$ is valued $\beta$ times the measured current $i$, for the example in Fig. \ref{fig:CCCS}, we add corresponding terms to KCLs for nodes $m$ and $n$. However, for the IM example specifically, we do not need an additional zero-valued voltage source as IM currents $(I_{qr}, I_{qs}, I_{dr},$ and $I_{ds})$ are already variables and easily accessible.

%\begin{wrapfigure}{L}{0.5\textwidth}
%\end{wrapfigure} 

\begin{figure}[htp]
    \centering
    \includegraphics[width=14cm]{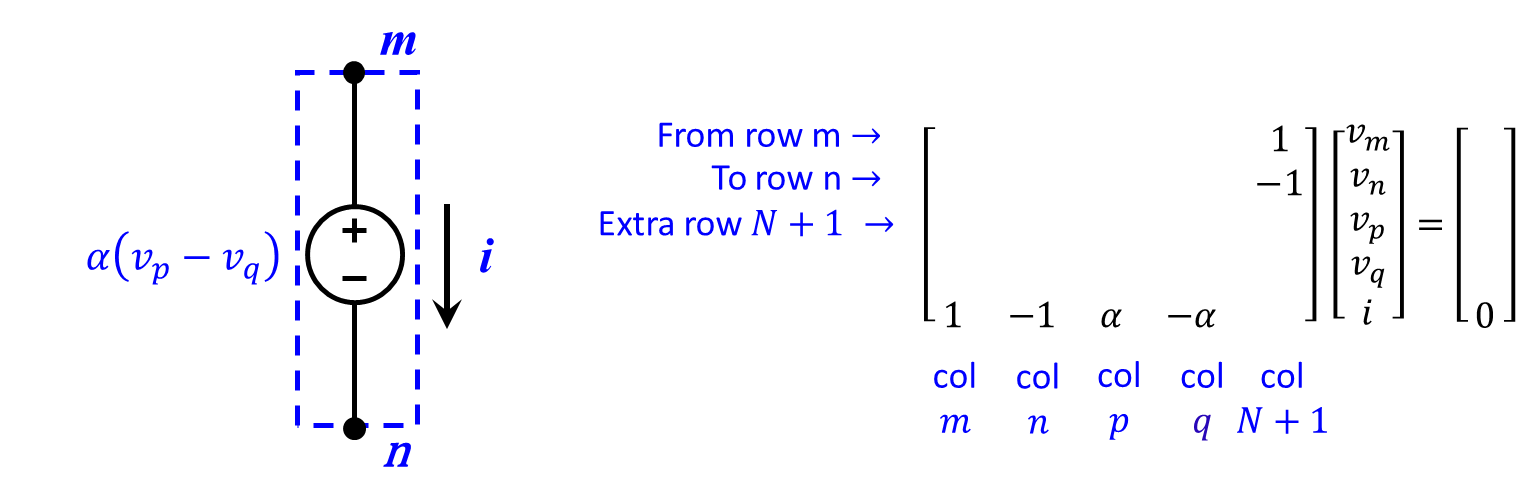}
    \caption{Stamps for VCVS. Reconstructed from Pileggi, Carnegie Mellon ECE 18-762 Notes \cite{circuit_simulation_pileggi}.}
    \label{fig:VCVS}
\end{figure}

The IM in dq frame is coupled to the rest of the circuit with controlled sources as well. These are given by $I^{IM}_a, I^{IM}_b, I^{IM}_c$ and $V_{ds}, V_{qs}$ in Fig. \ref{fig:IM_dq_load}. The first set $I^{IM}_a, I^{IM}_b, I^{IM}_c$, which map the currents consumed by the IM in $dq$ frame into $abc$ frame, are modeled via CCCS. The underlying math defining the relationship between currents in $dq$ and $abc$ frames is given by inverse $dq$ transformation. The stamps for these sources are covered in the CCCS description above. The second set, which maps the $abc$ voltages to $dq$ voltages with linear $dq$ transformation, is represented via voltage-controlled voltage sources (VCVS). The stamps for generic VCVS are shown in Fig. \ref{fig:VCVS}. Here, the voltage source across nodes $m$ and $n$ equals $\alpha$ times the difference between voltages at nodes $p$ and $q$. This relationship is captured in the extra node $N+1$ added to the matrix, which also gives us the current $i$ through the voltage source. The current is added to KCL rows for nodes $m$ and $n$. In the IM, the voltage $V_{ds}$ and $V_{qs}$ will be functions of $V_a^{IM}, V_b^{IM}$ and $V_c^{IM}$.

%\begin{wrapfigure}{L}{0.5\textwidth}
%\end{wrapfigure} 

\subsubsection{Stamping IM equations}

So far, we discussed addressing nonlinear and differential terms modularly for each component. We also learned how individual components are added to the system matrix. The next step is to learn how the various terms come together in the overall $Y$ matrix to satisfy system-level physics for the network in Fig \ref{fig:IM_dq_load}. We must also learn the updation of these terms for each iteration (for nonlinear elements) and each recursion (for each time step forward) to obtain a time-domain solution for the overall circuit. 

Let us begin with non-IM linear components in Fig \ref{fig:IM_dq_load}. We add them into the system matrix following the MNA approach described in Example 1 of this tutorial. We update the companion circuit terms for memory elements every instance we move forward in time. For the IM terms, we will use an alternative approach. If we were to stamp the IM circuit elements using MNA, we would end up with $\sim$ 20 additional nodes and corresponding equations for each IM we encounter in the network (refer to Fig. \ref{fig:IM_equi_circuit}). Therefore, to reduce the dimension of IM equations, in this tutorial, we instead used a combination of Kirchhoff Voltage Law (KVL)-based loop constraints and KCL-based nodal constraints to stamp the IM model's equivalent circuit elements. With this approach, we will stamp the equations for electrical sub-circuits (i through iv in Fig. \ref{fig:IM_equi_circuit}) in 4 new rows of the system matrix $Y$ following KVL-based loop constraints. Specifically, for each electrical sub-circuit in Fig. \ref{fig:IM_equi_circuit} and \eqref{eq:IM_i}-\eqref{eq:IM_iv}, moving from left to right, we will add stamps (for voltages) for each element such that the net sum of voltages in the loop is equal to zero. For the mechanical part (no. (v) in Fig. \ref{fig:IM_equi_circuit}), we will add terms to matrix $Y$ (including linearized $T_E$ and companion circuits $\frac{\partial \omega_r}{\partial t}$) following KCL-based nodal constraint. With this approach, we only add five new variables and corresponding rows for each instance of IM (instead of $\sim$20). The variables are $I_{ds}, I_{qs}, I_{dr}, I_{qr},$ and $\omega_r$.

\begin{figure}[htp]
    \centering
    \includegraphics[width=10cm]{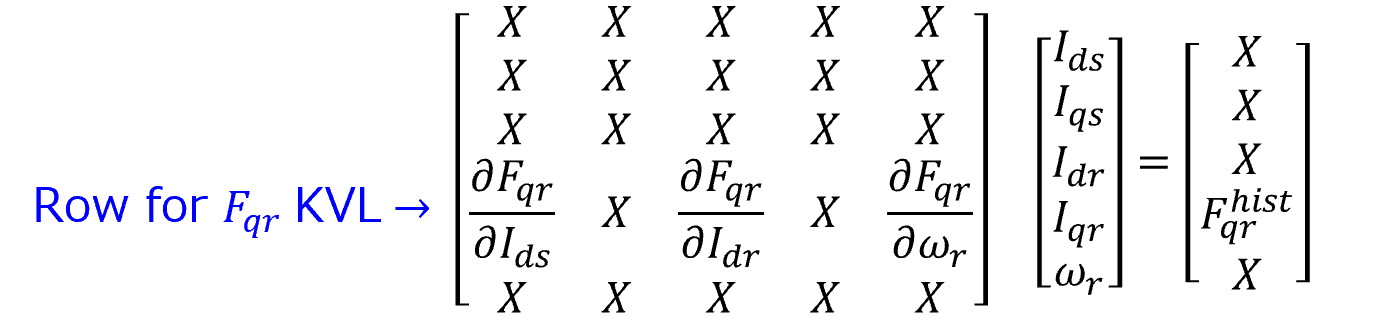}
    \caption{Nonlinear IM stamps for $F_{qr}$ KVL equation in \eqref{eq:IM_iv}.}
    \label{fig:IM_stamp_nonlinear}
\end{figure}

Recall that before stamping the terms into the system matrix $Y$, we have to perform two steps. First, we will replace the circuit elements with time-derivative terms with corresponding equivalent companion circuits. Second, we will replace the circuits with nonlinear terms with their linearized approximation as shown for $F_{qr}$ in  \eqref{eq:linearize_fqr}. For instance, the nonlinear symbolic stamps for $F_{qr}$ loop constraint in \eqref{eq:IM_iv} is added to the $Y$ matrix as shown in Fig. \ref{fig:IM_stamp_nonlinear}.

Next, to obtain the transient response of the nonlinear IM network, we will recursively solve the system matrix $Y$ over time, and we will perform iterations to solve the nonlinear terms. We will update the linearized terms in each iteration.  We will update the terms in companion circuits only when we recursively move forward in time.

\subsubsection{Initialization}

Next, we discuss how to initialize the IM network. In general, many approaches are available to initialize the network and we will discuss one such approach. 

\noindent Remember because of dq-transformation on IM variables, two sub-circuits evolve (see left and right of Fig. \ref{fig:IM_dq_load}). The sub-circuit on the left without the IM equations is a linear circuit. AC analysis can be used to obtain the initial condition for these, assuming a rated complex current draw by IM (which models the $I_a, I_b,$ and $I_c$ in Fig. \ref{fig:IM_dq_load}). A good guess for a complex current draw by IM can be obtained by running power flow with IM modeled as a PQ load and calculating the current from the solution ($I=S_{IM}^*/V_{IM}^*)$. Initialization of the IM components requires handling the nonlinearities due to speed-flux terms. With the proper choice of the reference frame in dq-transformation, time-invariant voltage sources ($V_{ds}$ and $V_{qs}$ are constants) across the IM can be obtained. We can then use DC analysis to obtain the initial conditions for the IM circuit. The $V_{ds},$ and $V_{qs}$ voltages are DC-values in the rotating reference frame. We can short the inductor and open the capacitors to obtain the steady-state initial conditions with the source voltages as DC values. The rated voltage at the IM terminals for dq-transformation can be obtained from the power flow solution $V_{IM}$. In reality, with a slightly more involved procedure, we can get exact initial conditions for the IM circuit by solving a set of nonlinear equations representing the overall circuit iteratively using NR.

Generally, one must note that initializing large complex EMT networks is not trivial, and many commercial tool manuals \cite{pscad_initializer_manual}, book chapters \cite{circuit_simulation_pileggi}, and research papers are devoted to the study of efficiently initializing the network. In power systems, a common practice is to run load flow and then map the load flow solution from the positive sequence frequency domain to three time-domain to initialize large power grid EMT networks. However, with emerging inverter-based resources, this approach may no longer work. 

\subsubsection{Nonlinear time-domain simulation}

\begin{figure}[H]
    \centering
    \includegraphics[width=14cm]{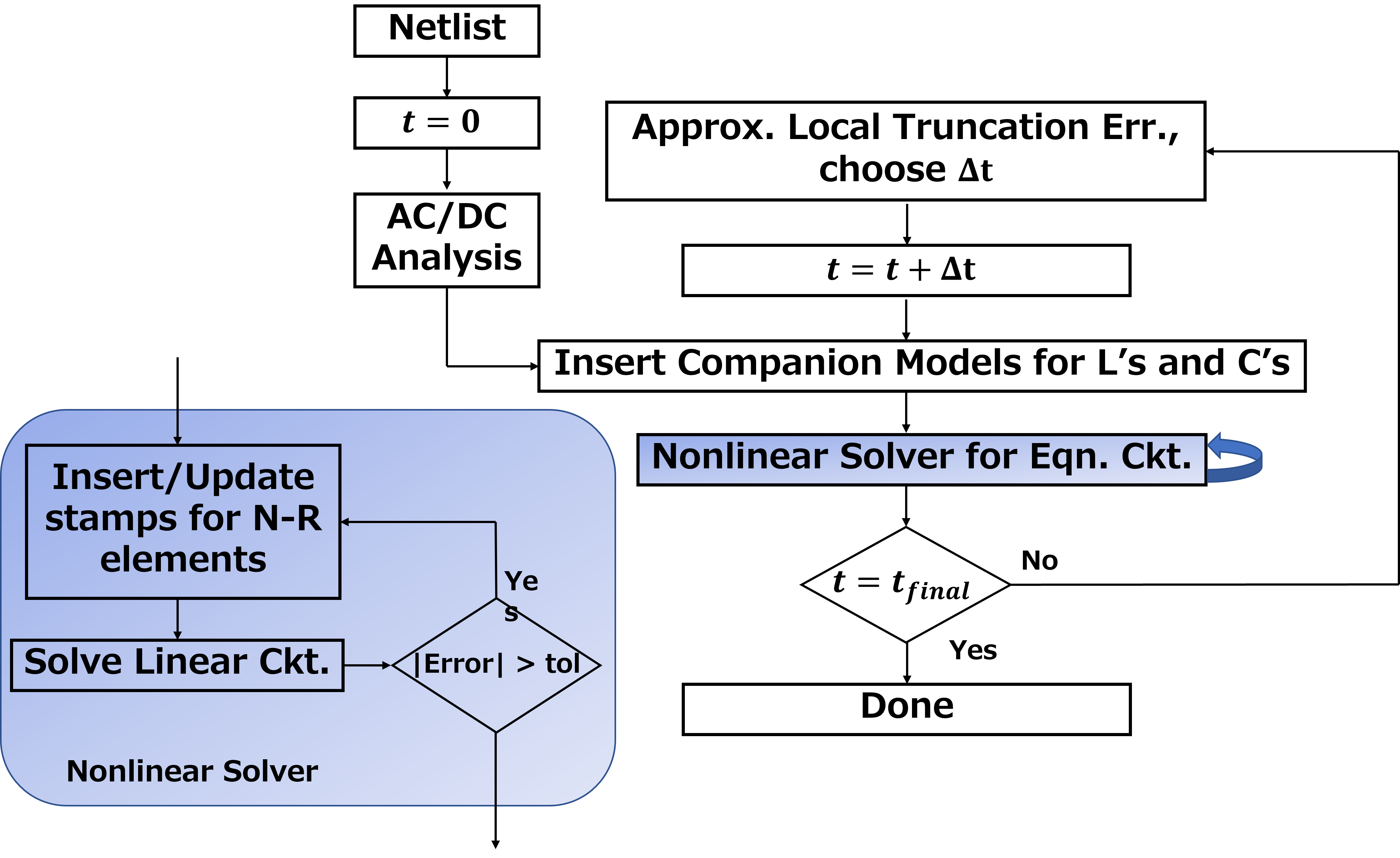}
    \caption{Running EMT simulation for nonlinear networks.}
    \label{fig:nonlinear_analysis_flowchart}
\end{figure}

For the nonlinear IM problem, with access to the initial state at $t=0$, we repeatedly solve a system matrix $Y$ (moving forward by $\Delta t$ in each recursion) following steps described in Fig. \ref{fig:nonlinear_analysis_flowchart} to obtain the time-domain response from $t=0$ to $t=t_{final}$. How we update the system matrix in the nonlinear analysis is different than in the case of linear analysis in Section \ref{sec:linear_analysis}. Linearized terms in the system matrix $Y$ corresponding to nonlinear devices are updated \textit{iteratively} after each linear solve. The memory-stamps corresponding to the devices  with differential terms are only updated each time we move forward in time (\textit{recursion}). We only move forward in time once the nonlinear stamps have converged (i.e., no considerable difference in the value of linearized stamps in subsequent iterations). As in the linear case, we dynamically adjust $\Delta t$ at each step based on the trade-off between simulation run-time and local truncation error (LTE).

\section*{Conclusion and Resources for a Deeper Dive}

This document covers the fundamentals of building power systems EMT tools using circuit theory with the aid of two simple examples. The tutorial is introductory, and it does not include many critical topics in detail, such as solver initialization, circuit-simulation heuristics for steep nonlinear models, sparse matrices, new power electronics-based models, and model reduction. For a deeper dive into the subject of EMT simulations, I recommend the following references \cite{dommel1996emtp}, \cite{dommel1969digital}, \cite{circuit_simulation_pileggi}, \cite{pscad_initializer_manual}, and \cite{milano2010power}.

\section*{Sample Parameters}
Parameter values for the two examples and a few others are included in the .json format and shared in the following publicly available git repo: \url{https://github.com/amritanshup7/Tutorial-Circuit-based-Electromagnetic-Transient-Simulation/blob/main/example2_IM/IM_circuit.json}. Anyone interested in EMT tool implementation for the two examples in Python language can reach out to me at \url{amritanshu.pandey@uvm.edu}.

\section{Acknowledgments}
Tim McNamara and Naeem Turner-Bandele have both TA'ed the course where this material was taught and they have spent a significant amount of time ensuring that the derivations and explanations in this document are precise and accurate. Tim also provided valuable feedback on the document's structure and content.

I also want to acknowledge my Ph.D. advisor Larry Pileggi whose book, lecture slides, and guidance were critical in compiling this document.

\printbibliography

\end{document}